\documentclass[11pt]{article}

\usepackage{amsmath,amssymb}
\usepackage{slashed}
\usepackage{xcolor} 
\usepackage{graphicx}
\usepackage{dcolumn} 
\usepackage{bm} 
\usepackage{epsfig}
\usepackage{epstopdf}
\usepackage{grffile}
\usepackage{color}
\usepackage{colordvi}
\usepackage{amsmath,amssymb}
\usepackage{rotating}
\usepackage{lscape}
\usepackage{cite}
\usepackage{float}
\usepackage{hyperref}
\usepackage{url}

\usepackage{amsmath}
\usepackage{amssymb}
\usepackage{caption}

\newcommand{\delphes}{{\sc Delphes}}
\newcommand{\madgraph}{{\sc MadGraph}}

\newcommand{\feynrules}{{\sc Feyn\-Rules}}
\newcommand{\pythia}{{\sc Pythia}}

\newcommand{\beq}{\begin{equation}}
\newcommand{\eeq}{\end{equation}}

\newcommand{\met}{\slashed{p}_T}
\newcommand{\cO}{\mathcal{O}}
\newcommand{\cL}{\mathcal{L}}

\def\Re{{\cal R \mskip-4mu \lower.1ex \hbox{\it e}\,}}
\def\Im{{\cal I \mskip-5mu \lower.1ex \hbox{\it m}\,}}

\def\tev{\,{\ifmmode\mathrm {TeV}\else TeV\fi}}
\def\gev{\,{\ifmmode\mathrm {GeV}\else GeV\fi}}
\def\mev{\,{\ifmmode\mathrm {MeV}\else MeV\fi}}

\begin{document}

\begin{center}

\vspace*{15mm}
\vspace{1cm}
{\Large \bf Probing FCNC couplings in single top quark production associated with a neutral gauge boson at future lepton colliders}

\vspace{1cm}

\small{\bf Sara Khatibi$^{\dagger}$\footnote{sara.khatibi@ut.ac.ir},  
Mehrnoosh Moallemi$^{\ddagger}$\footnote{mehrnoosh.moalemi@gmail.com} }

 \vspace*{0.5cm}

{\small\sl 
$^{\dagger}$Department of Physics, University of Tehran, North Karegar Ave., Tehran 14395-547, Iran\\
$^{\ddagger}$Department of Physics, Isfahan University of Technology, Isfahan 84156-83111, Iran \\
}

\vspace*{.2cm}
\end{center}

\vspace*{10mm}

%

%
%
\begin{abstract}\label{abstract}
In this paper, we consider a single top quark production in association with a neutral gauge boson
(Z boson or Photon) at the future electron-positron colliders. We utilize these channels
to probe the top quark Flavour Changing Neutral Currents (FCNC) interactions in $tqZ$ and $tq\gamma$ 
vertices. We perform two separate analyses for top-Z-jet and  top-$\gamma$-jet channels. 
The Standard Model Effective Field Theory (SMEFT) approach is employed to search for these anomalous couplings.
We consider parton showering, hadronization and fast detector simulation in our study and use a cut-and-count 
technique to separate the signal from the Standard Model (SM) background processes.
The upper limits on the FCNC couplings at $95\%$ confidence level for the different 
integrated luminosities are obtained. It is shown that the future lepton collider would be able to probe the FCNC
branching fractions to $\rm{Br}(t\rightarrow q \gamma)< 10^{-4}$ and $\rm{Br}(t\rightarrow q Z) < 10^{-4}$
with $3$ $\rm{ab}^{-1}$of integrated luminosity of data at a center of mass energy of $350$~GeV.
\end{abstract}

\newpage


\section{Introduction}
\label{intro}
The Standard Model (SM) of particle physics has been found to be a successful theory 
in describing particles and their interactions at~$\tev$ scale.
So far, all experimental results are in a good agreement with the SM predictions.
However, the SM fails to answer some major experimental observations like the need 
for the existence of Dark Matter (DM) in the cosmos, 
the massive neutrinos, baryon asymmetry in the universe, etc. 
Hence, particle physicists have looked for new physics for several decades.

Despite all the plentiful efforts at the Large Hadron Collider (LHC) at run I and run II, 
no signal of new physics has been found yet.
Therefore, it seems that the new physics scale should be well separated from the electroweak scale
and the new degrees of freedom cannot be produced directly at the colliders. 
So, in order to study the new physics, we should look for indirect signs such as new interactions 
between the SM particles. 
Flavour changing neutral current (FCNC) interactions are a good place to seek for these new effects since
they are forbidden at the tree level in the SM framework and they are also strongly suppressed at the loop level 
by the Glashow-Iliopoulos-Maiani (GIM) mechanism~\cite{Glashow:1970gm}. 

Particularly, the search for FCNC interactions in the top sector is well-motivated, 
because of the especial properties of top quark. 
The SM predicts very small branching ratios ($\cO(10^{-14})$) for 
different top quark FCNC decay modes~\cite{Agashe:2013hma}, 
hence the current and future experiments would not be able to probe such tiny values.
Nevertheless, there are many Beyond Standard Models (BSM) such as 
Technicolor, Minimal Supersymmetric Standard Model (MSSM), 
two Higgs Doublet Models (2HDM) that anticipate for large branching ratios 
for the top FCNC decay modes~\cite{Luke:1993cy,Atwood:1996vj,Liu:2004qw,Delepine:2004hr,Li:1993mg,Lopez:1997xv,Yang:1997dk}.
Therefore, any footprint of the top FCNC interactions can prove the existence of the new physics.

There are several phenomenological papers that studied the top FCNC couplings at different 
colliders in the literature~\cite{AguilarSaavedra:2004wm,AguilarSaavedra:2000aj,AguilarSaavedra:2002ns,Khatibi:2015aal,Khanpour:2014xla,Khatibi:2014via,Gao:2011fx,Khanpour:2019qnw,Hesari:2015oya,Hesari:2014eua,Etesami:2010ig}.
Furthermore, there are many experimental analyses that look for the top FCNC interactions using different channels~\cite{Aad:2015uza,Aad:2015gea,Aad:2012ij,Khachatryan:2015att,Aad:2019pxo,Aaboud:2018nyl,Chatrchyan:2012hqa,Chatrchyan:2013nwa,Abazov:2011qf,Abramowicz:2011tv,Abdallah:2003wf,Sirunyan:2017kkr}.
For instance, the CMS collaboration has studied the top quark FCNC interactions 
in the anomalous single top quark production in association with a photon 
in proton-proton collision at 8 TeV~\cite{Khachatryan:2015att}.
Following upper limits at the $95\%$ CL on the anomalous top branching ratios 
have been found; $\rm{Br}(t\rightarrow u \gamma)<1.3 \times 10^{-4}$ and 
$\rm{Br}(t\rightarrow c \gamma)<1.7 \times 10^{-3}$.
Recently, more stringent bounds on these branching ratios 
have been found by ATLAS detector at a center of mass energy of 13 TeV;  
$\rm{Br}(t\rightarrow u \gamma)< 6.1 \times 10^{-5}$ and 
$\rm{Br}(t\rightarrow c \gamma)< 18 \times 10^{-5}$~\cite{Aad:2019pxo}.
In addition, the ATLAS experiment has searched for the FCNC couplings 
in proton-proton collision at 13 TeV by using top-quark pair events which one
of them decaying through $t \rightarrow q Z$ and another one decays through 
the standard mode~\cite{Aaboud:2018nyl}.
The collaboration has found the upper limits at the $95\%$ CL on the top branching 
ratios as $\rm{Br}(t\rightarrow u Z)<1.7 \times 10^{-4}$ and 
$\rm{Br}(t\rightarrow c Z)<2.4 \times 10^{-4}$.
Some papers studied the FCNC couplings in the HL-LHC, the reported values for branching ratios are
$\rm{Br}(t\rightarrow u \gamma)<4.6 \times 10^{-5}$, $\rm{Br}(t\rightarrow c \gamma)<3.4 \times 10^{-4}$, $\rm{Br}(t\rightarrow u Z )(\sigma_{\mu\nu}) <8.1 \times 10^{-4}$
and $\rm{Br}(t\rightarrow u Z )(\gamma_{\mu})<1.7 \times 10^{-3}$~\cite{Malekhosseini:2018fgp, CMS-DP-2016-064,CMS-PAS-FTR-13-016}.

There are several proposed future electron-positron colliders for very 
precise measurements such 
as the future circular electron–positron collider (FCC-ee)~\cite{Abada:2019lih,Abada:2019zxq}, 
the Circular electron-positron Collider (CEPC)~\cite{CEPC-SPPCStudyGroup:2015csa}, 
the International Linear Collider (ILC)~\cite{Aihara:2019gcq,Baer:2013cma,Behnke:2013lya}, 
and the Compact Linear Collider (CLIC)~\cite{Abramowicz:2013tzc}.
Since large amounts of production of top quarks would happen at these colliders, 
the top quark couplings can be probed at high precision. 
Our main purpose here is to study the top FCNC couplings 
in the single top quark production in association with a
neutral gauge boson (Z or $\gamma$) at a future electron-positron collider. 
This analysis is a complementary channel in addition to other channels to probe 
the anomalous FCNC couplings, since the top-Z/$\gamma$-jet has a different signature at the lepton colliders.

In this paper, these signals are studied separately. 
In the top-Z-jet analysis, we consider both top quark and Z boson decay leptonically, 
therefore, we expect to have three charged leptons, a b-jet, a light-jet, 
and the missing energy in the final state.
However, in the top-$\gamma$-jet analysis, leptonic decay for the top quark is assumed, 
so the final state comprises a charged lepton, a hard photon, a b-jet, a light-jet, 
and the missing energy due to the neutrino.
Furthermore, parton showering and hadronization as well as the detector simulations 
are taken into account in these studies. 
Then, we determine the upper limits on the top FCNC couplings and the corresponding branching 
ratios in the both analyses.

The paper is structured as follows. In section~\ref{sec:model}, we introduce the 
effective Lagrangian which describes the top quark FCNC interactions with the neutral gauge bosons.
In section~\ref{sec:tZjet}, we explain our analysis strategy like event 
simulations and event selections for top-Z-jet signal 
and then present our obtained limits on the new couplings and the corresponding branching ratios.
Analysis strategy for top-$\gamma$-jet signal and the related upper limits on the anomalous couplings 
are described in section~\ref{sec:tAjet}.
Finally, we present our summary in section~\ref{sec:summary}.

\section{Theoretical formalism}
\label{sec:model}
In this section, we explain the theoretical framework which we use to search for FCNC couplings.
There are two different approaches to study any new physics at the colliders. 
The first approach is the direct searches for the well-motivated BSM models. 
In another word, we look for the new degrees of freedom which anticipated by these new models at the colliders.
The second approach is the indirect searches which mean we seek for the new interactions between SM particles.  

Since all the experimental measurements are in good agreement with the  SM predictions, 
the new physics scales seem to be well separated from the SM scale.
On the other hand, because of the presence of several new physics scenarios and sometimes 
with the same signatures at the colliders, an economical way to search for the new physics would be using a
model independent way. The Standard Model Effective Field Theory (SMEFT) approach 
provides this model independent way to look for any sign of new physics~\cite{Brivio:2017vri}. 

In the SMEFT, all the new degrees of freedom are integrated out and the new physics effects show up
in the higher dimensional effective operators which contain the new interactions between the SM field content.
These higher dimensional operators are suppressed by the inverse powers of the scale of new physics and 
respect the Lorentz and SM gauge symmetries. 
The operators with odd dimensions are not considered by assuming baryon and lepton number conservation.
Furthermore, the EFT approach is useful when the scale of new physics is much larger than the 
energy scale of the process.
Since the operators are suppressed by the powers of $\Lambda$, 
one expects that the smallest dimension provides the leading corrections to the SM amplitudes.
The effects of dimension eight and higher-order operators are suppressed by at least $1/\Lambda^4$. 
So, the dimension six operators have dominant contributions to the observables at the colliders. 
Worth mentioning that, if the coefficients of dimension six operators, which can contribute to the special process, are strongly constrained by current experiments, the dimension eight operators have a main contribution to the process.
The SMEFT Lagrangian up to dimension six operators could be written as following~\cite{Grzadkowski:2010es}:
\begin{equation}
\mathcal{L}_{\rm{SMEFT}}=\mathcal{L}_{\rm{SM}}+\sum_{i} \frac{C_{i} O_{i}}{\Lambda^{2}},
\end{equation}
here the first term is the SM Lagrangian with four dimensional operators and $O_{i}$ demonstrates operators 
with dimension six. The new physics scale is denoted by $\Lambda$ and $C_{i}$ are the Wilson couplings.
The following dimension six operators can contribute in the FCNC interactions of a top quark and a Z boson/photon 
with an up-type quark (u or c quark)~\cite{AguilarSaavedra:2008zc}:
\begin{eqnarray}
O_{u W}^{i j}&=&\left(\bar{q}_{L i} \sigma^{\mu \nu} \tau^{I} u_{R j}\right) \tilde{\phi} W_{\mu \nu}^{I}, \nonumber\\
O_{u B \phi}^{i j}&=&\left(\bar{q}_{L i} \sigma^{\mu \nu} u_{R j}\right) \tilde{\phi} B_{\mu \nu},\nonumber\\
O_{\phi q}^{(3, i j)}&=&i\left(\phi^{\dagger} \tau^{I} D_{\mu} \phi\right)\left(\bar{q}_{L i} \gamma^{\mu} \tau^{I} q_{L j}\right),\nonumber\\
O_{\phi q}^{(1, i j)}&=&i\left(\phi^{\dagger} D_{\mu} \phi\right)\left(\bar{q}_{L i} \gamma^{\mu} q_{L j}\right),\nonumber\\
O_{\phi u}^{i j}&=&i\left(\phi^{\dagger} D_{\mu} \phi\right)\left(\bar{u}_{R i} \gamma^{\mu} u_{R j}\right),
\end{eqnarray}
where $q_{L i}$ and $u_{R j}$ are left-handed quark doublet and right-handed quark singlet, respectively, 
and $i , j$ are flavour indices. The Pauli matrices are presented with $\tau^{I}$ and $D_{\mu}$ is covariant derivative.
Moreover, $\tilde{\phi}=i \tau^2 \phi^{*}$ that $\phi$ is the SM Higgs doublet.
Their contribution to the FCNC interactions can be parametrized as follows,
\begin{align}
\mathcal{L}_{\rm{FCNC}} = \sum_{q=u, c} \lbrace &-\frac{g_{W}}{2 c_{W}} \bar{q} \gamma^{\mu}\left(X_{tq}^{L} P_{L}+X_{tq}^{R} P_{R}\right) t Z_{\mu}+\frac{g_{W}}{4 c_{W} m_{Z}} \bar{q} \sigma^{\mu \nu}\left(\kappa_{t q}^{L} P_{L}+\kappa_{t q}^{R} P_{R}\right) t Z_{\mu \nu}  \nonumber \\
\quad &+ \frac{e}{2 m_{t}} \bar{q} \sigma^{\mu \nu}\left(\lambda_{t q}^{L} P_{L}+\lambda_{t q}^{R} P_{R}\right) t A_{\mu \nu}\rbrace+ h.c.  ,
\label{eq:FCNC-lag}
\end{align}

where the field strengths of the Z boson and photon are shown by $Z_{\mu \nu}$ and $A_{\mu \nu}$, respectively.
Also, $P_{R}$ and $P_{L}$ demonstrate the right and left-handed projection operators, respectively.
The dimensionless anomalous couplings between a top quark, a light up-type quark and a Z boson are presented by
$X_{t q}^{L}$, $X_{t q}^{R}$, $\kappa_{t q}^{L}$ and $\kappa_{t q}^{R}$. 
Similarly, $\lambda_{t q}^{L}$ and $\lambda_{t q}^{R}$ are dimensionless anomalous couplings between a 
top quark, a light up-type quark and a photon. These dimensionless anomalous couplings have the following 
relation with the Wilson coefficients of effective operators~\cite{AguilarSaavedra:2008zc}, 
\begin{eqnarray}
 X_{tq}^{L}&=&\frac{1}{2}\left[C_{\phi q}^{(3,j3)}+C_{\phi q}^{(3,3j) *}-C_{\phi q}^{(1,j3)}-C_{\phi q}^{(1,3j) *}\right] \frac{v^{2}}{\Lambda^{2}}, \nonumber\\
 X_{tq}^{R}&=&-\frac{1}{2}\left[C_{\phi u}^{j3}+C_{\phi u}^{3j *}\right] \frac{v^{2}}{\Lambda^{2}}, \nonumber\\
 \kappa_{tq}^{L}&=&\sqrt{2}\left[c_{W} C_{u W}^{3j *}-s_{W} C_{u B \phi}^{3j *}\right] \frac{v^{2}}{\Lambda^{2}},\nonumber \\
\kappa_{tq}^{R}&=&\sqrt{2}\left[c_{W} C_{u W}^{j3}-s_{W} C_{u B \phi}^{j3}\right] \frac{v^{2}}{\Lambda^{2}},\nonumber\\
 \lambda_{tq}^{L} &=&\frac{\sqrt{2}}{e}\left[s_{W} C_{u W}^{3j *}+c_{W} C_{u B \phi}^{3j*}\right] \frac{v m_{t}}{\Lambda^{2}}, \nonumber\\
\lambda_{tq}^{R} &=&\frac{\sqrt{2}}{e}\left[s_{W} C_{u W}^{j3}+c_{W} C_{u B \phi}^{j3}\right] \frac{v m_{t}}{\Lambda^{2}} ,
\label{Eq:Couplings}
\end{eqnarray}
where $j$ can be $1$ and $2$ corresponding to u-quark and c-quark couplings, respectively.

\begin{figure*}[t]
	\begin{center}
		\includegraphics[width=1\textwidth]{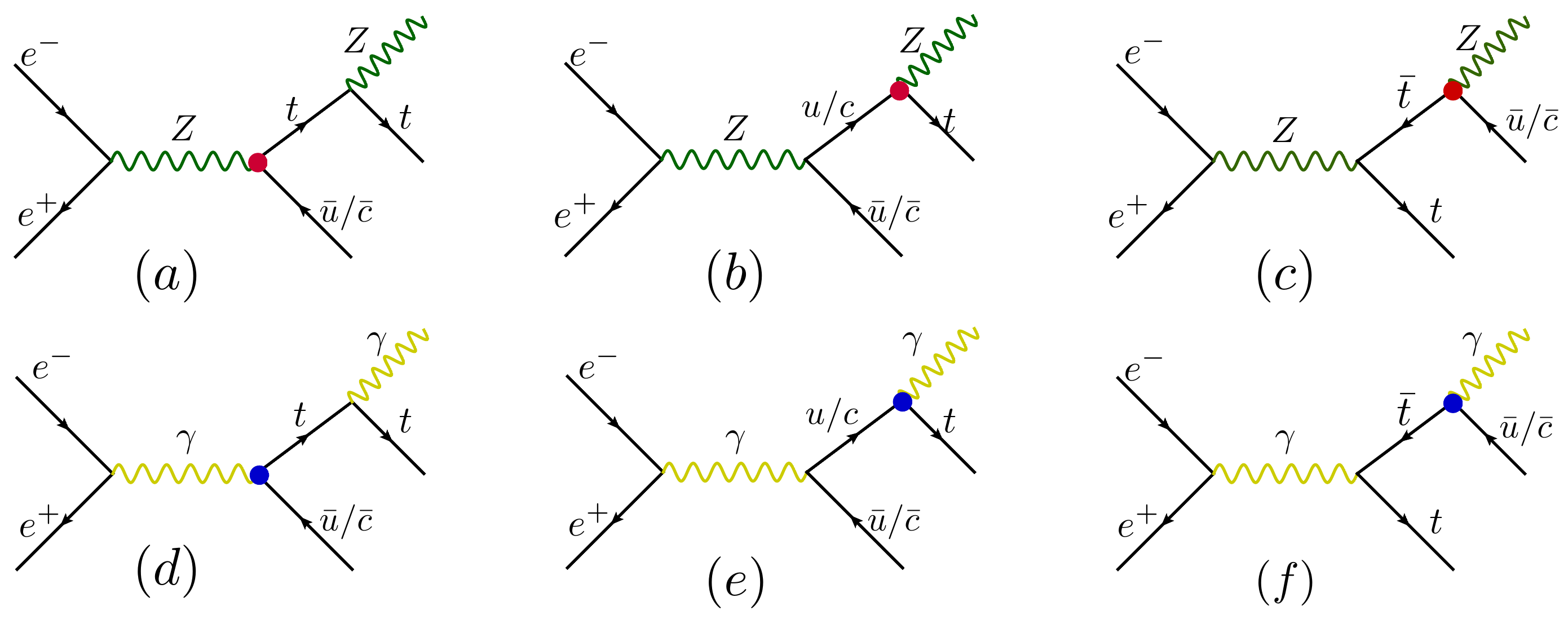}
		\caption{Representative Feynman diagrams for the single top quark production in association 
		with a neutral gauge boson in the presence of the FCNC couplings are shown. 
		Diagrams $a$, $b$, and $c$ arise from the anomalous couplings $tZq$ ($X_{tq}$ or $\kappa_{tq}$) 
		and diagrams $d$, $e$, and $f$ occur in presence of the anomalous couplings $t\gamma q$ ($\lambda_{tq}$).}
		\label{Feynman}
	\end{center}
\end{figure*}

As mentioned before, we are interested to look for the single top production in association with 
a Z boson or a photon and a light jet at a future electron-positron collider. 
This process cannot be produced in the SM framework, however, above FCNC interaction terms allow to generate 
such processes. Some Feynman diagrams of these processes in presence of the FCNC couplings 
are illustrated in Fig.~\ref{Feynman}. Diagrams $a$, $b$, and $c$ 
arise from the anomalous couplings $tqZ$ ($X_{t q}$ or $\kappa_{t q}$) and diagrams $d$, $e$, and $f$ occur
in presence of the anomalous couplings $tq\gamma$ ($\lambda_{t q}$).
We should mention that the $e^+e^- \rightarrow t \bar{t} \rightarrow t V j (V=Z~\text{or}~\gamma)$ process
have contribution in our final state and  we considered it in our analysis as well (diagrams $c$ and $f$).

In the two next sections, we use the effective Lagrangian (Eq.~\ref{eq:FCNC-lag}) to generate and analyse events for the 
top quark production with a neutral gauge boson signal to study FCNC interactions. 
We explain extensively analysis for the top-Z-jet signal in section~\ref{sec:tZjet} 
and analysis for the top-$\gamma$-jet signal in~\ref{sec:tAjet}.

\section{Analysis Strategy for top-Z-jet signal}
\label{sec:tZjet}

In this section, we study the single top quark production in association with a Z boson to look for 
the $tqZ$ and $tq\gamma$ anomalous couplings at the future electron-positron collider.
In the first subsection, event generation and simulation for both the signal and backgrounds are explained in detail.
Then, we describe our reconstruction method of the final state particles and 
introduce some strategies to reduce the number of background events with respect to the number of signal events. 
Eventually, the upper limit which can be extracted from the top-Z-jet signal is presented.

\subsection{Event Generation and Simulation}
\label{sec:EG-tZjet}

Now we are ready to define the top-Z-jet signal and related background processes 
and describe the event generation and simulation method.
Since the mass of final state particles, the top quark and Z boson, 
are heavy to be produced in the collider with $\sqrt {s} = 240~\rm {GeV}$, 
we just focus on the center of mass energy of $\sqrt {s} = 350~\rm {GeV}$ in this analysis. 
The signal process is single top production in association with a 
Z boson and a light jet ($e^-~e^+ \rightarrow t/\bar{t}~Z~j$), 
which both top quark and Z boson decay leptonically. 
As a result, the final state consists of three charged leptons (electron or muon), a b-jet, a light jet, 
and the missing energy. 
The representative Feynman diagrams of the signal production are presented in Fig.~\ref{Feynman} ($a$, $b$ and $c$).
Regarding generate of the signal events, we insert the UFO module~\cite{Degrande:2011ua} 
in the \madgraph~5 package~\cite{Alwall:2014hca}.
This UFO module has been obtained from implementing the
effective Lagrangian of Eq.~\ref{eq:FCNC-lag}~\cite{Amorim:12} 
in the \feynrules~package~\cite{Alloul:2013bka}.

Since all the $tqZ$ and $tq\gamma$ FCNC interactions produce our favorable signal (top-Z-jet),  
the signal events are categorized into three scenarios.
 In the first scenario, we choose a non zero value for $\kappa_{tq}^{R}=\kappa_{tq}^{L}=\kappa_{tq}$ 
 and put zero for other anomalous couplings. 
 In the second one, we put zero value for all FCNC couplings except $X_{tq}^{R}=X_{tq}^{L}=X_{tq}$. 
 And in the final case, just $\lambda_{tq}^{R}=\lambda_{tq}^{L}=\lambda_{tq}$ coupling has a non zero value.
Also, it should be mentioned that we set the same values for the u-quark and c-quark anomalous couplings.

According to the final state of the signal, the main background processes are
$W^{+}W^{-}jj$, $W^{\pm}Zjj$, $t\bar{t}$, and $ZZjj$.
In this processes all the W-boson, Z-boson, and top quark decay leptonically.
Since tau lepton can have a hadronic decay, it indicates a different signature in the detector, 
so decay to electron and muon are just taken into account.
The \madgraph~5 package is used to generate all the background processes.
At generator level,  preselection cuts like $p_T \geq 10$ GeV and $|\eta|< 3$ are applied on all objects in
the final state.\footnote{ For applying cuts we use $p_T$ and $\eta$, however, we should mention that
at sub-TeV lepton colliders, energy and angle are much more standard variables. }
The angular distance between all the charged leptons and jets has to 
be greater than $0.4$ ($\Delta R=\sqrt{(\Delta \eta)^{2}+(\Delta \phi)^{2}}$ ).
For both signal and backgrounds, the \pythia~package~\cite{Skands:2014pea} 
is employed for parton showering and hadronization. 
Then, the \delphes~package~\cite{deFavereau:2013fsa} is utilized to model the detector performance 
with using the ILD card~\cite{Baer:2013cma}.
For numerical calculation, we use following the SM inputs: $m_{\text{top}}=173.3$ GeV,
$m_{Z}=91.187$ GeV,
$m_{W}=80.419$ GeV,
$\alpha=1/127.90$, and
$s^2_{W}=0.234$~\cite{Tanabashi:2018oca}.

\begin{table}[t]
\begin{center}
\begin{footnotesize}
\begin{tabular}{|c|ccc|cccc|}
\hline
Cross-sections (in fb)     &  \multicolumn{3}{|c|}{ Signal }       &\multicolumn{4}{|c|}{ Background }    \\ 
\hline
Cuts     &   $tqZ(\sigma_{\mu\nu})$   &    $tqZ(\gamma_{\mu})$  & $tq\gamma $  &  $WWjj$&$WZjj$& $t\bar{t}$ & $ZZjj$  \\
\hline  \hline
Preselection cuts     & $13.5(\kappa_{tq})^{2}$ &   $9.28(X_{tq})^{2}$  &  $0.14(\lambda_{tq})^{2}$   & $10.12$  &  $0.26$  &   $9.25$  &   $0.007$\\
$1~\rm{light~jet} + 1\rm{b~jet}$    & $6.92(\kappa_{tq})^{2}$ &   $4.76(X_{tq})^{2}$     &  $0.066(\lambda_{tq})^{2}$  & $4.43$  &  $0.012$   &   $4.2$ &   $0.001$  \\
$3~\rm{leptons}$    & $3.55(\kappa_{tq})^{2}$ &   $2.5(X_{tq})^{2}$   &  $0.03(\lambda_{tq})^{2}$  &  $0.00014$ &   $0.006$  &   $0.00024$   &   $0.0007$  \\
$ \met >30~\rm{GeV}$       & $2.6(\kappa_{tq})^{2}$  &   $1.8(X_{tq})^{2}$    &  $0.02(\lambda_{tq})^{2}$   &  $0.00012$  &   $0.004$  &   $0.00022$   &   $0.00001$  \\ 
\hline
\end{tabular}
\end{footnotesize}
\caption{Cross sections for the  top-Z-jet signal and the SM background processes before and after passing sequential preliminary cuts at the center of mass energy of $\sqrt {s} $ = 350 GeV.}
\label{firstCuts-tqz}
\end{center}
\end{table} 
  
In order to select the signal events, we desire to have at least two jets with $p_{T} \geq 20$ GeV and $|\eta| < 2.4$, 
which at least one of them is required to be tagged as a b-jet.
We require the events with at least three charged leptons with $p_{T} \geq 10$ GeV and $|\eta| < 2.4$. 
Furthermore, we choose events with the missing
transverse momentum greater than $30$ GeV. 
The cross sections including the branching ratios for three signal scenarios 
and the corresponding SM backgrounds after passing sequential preliminary cuts 
at the center of mass energy of 350 GeV are presented in Table~\ref{firstCuts-tqz}.
It is worth mentioning that the reported cross sections for signal and background processes 
are not included the QED ISR.
Moreover, it's noteworthy that the gamma-induced processes or e-gamma processes can be 
the new sources of background, however, we expect that by applying our preliminary 
and secondary cuts their contribution will be strongly suppressed. So, we don't consider them in our analysis.

As it is clear, the cross section of the signal for $tq\gamma$ scenario is significantly less than two others, 
so we expect that the extracting upper limit for this coupling should be looser than two others.
Also, the cross section of the signal for $tqZ(\sigma_{\mu\nu})$ scenario is greater 
than $tqZ(\gamma_{\mu})$, because of the presence of the gauge boson's momentum in the 
$tqZ(\sigma_{\mu\nu})$ interaction term. The preliminary cuts efficiencies 
for $tqZ(\sigma_{\mu\nu})$, $tqZ(\gamma_{\mu})$, and $tq\gamma$ signal scenarios 
are $0.19$, $0.19$,  and $0.14$,  respectively.

It is notable that, although the production cross sections for some background processes like $WWjj$ and $t\bar{t}$ 
are large at parton level, they are remarkably reduced after applying preliminary cuts, particularly,
when we require exactly three charged leptons at the final state. 
The preliminary cuts efficiencies for $WWjj$, $WZjj$,  $t\bar{t}$, and $ZZjj$ backgrounds
are at the order of $10^{-5}$, $10^{-2}$,  $10^{-4}$, and $10^{-3}$, respectively. Therefore, our main backgrounds 
are $WZjj$ and $ZZjj$ channels in this analysis.

\subsection{Separation of Signal from Background}
\label{sec:SBtZj}
For suppressing the contributions of background events, we look at different
kinematic distributions to find suitable secondary cuts.  
Therefore, we need to reconstruct the top quark and Z boson of the final state.
In order to reconstruct the Z boson, we require two opposite charged leptons
with an invariant mass between 60 to 120 GeV. 
Then the top quark will be reconstructed with the remaining charged lepton, the b-jet and the neutrino.
As it is clear from Table~\ref{firstCuts-tqz} that the signal cross sections for 
$\kappa_{tq}$ and $X_{tq}$ couplings are larger than the signal cross section for $\lambda_{tq}$,
so we first describe our secondary cuts for $tqZ$ interactions ($\sigma_{\mu\nu}$ and $\gamma_{\mu}$).

\begin{figure*}[t]
	\begin{center}
		\includegraphics[width=0.45\textwidth]{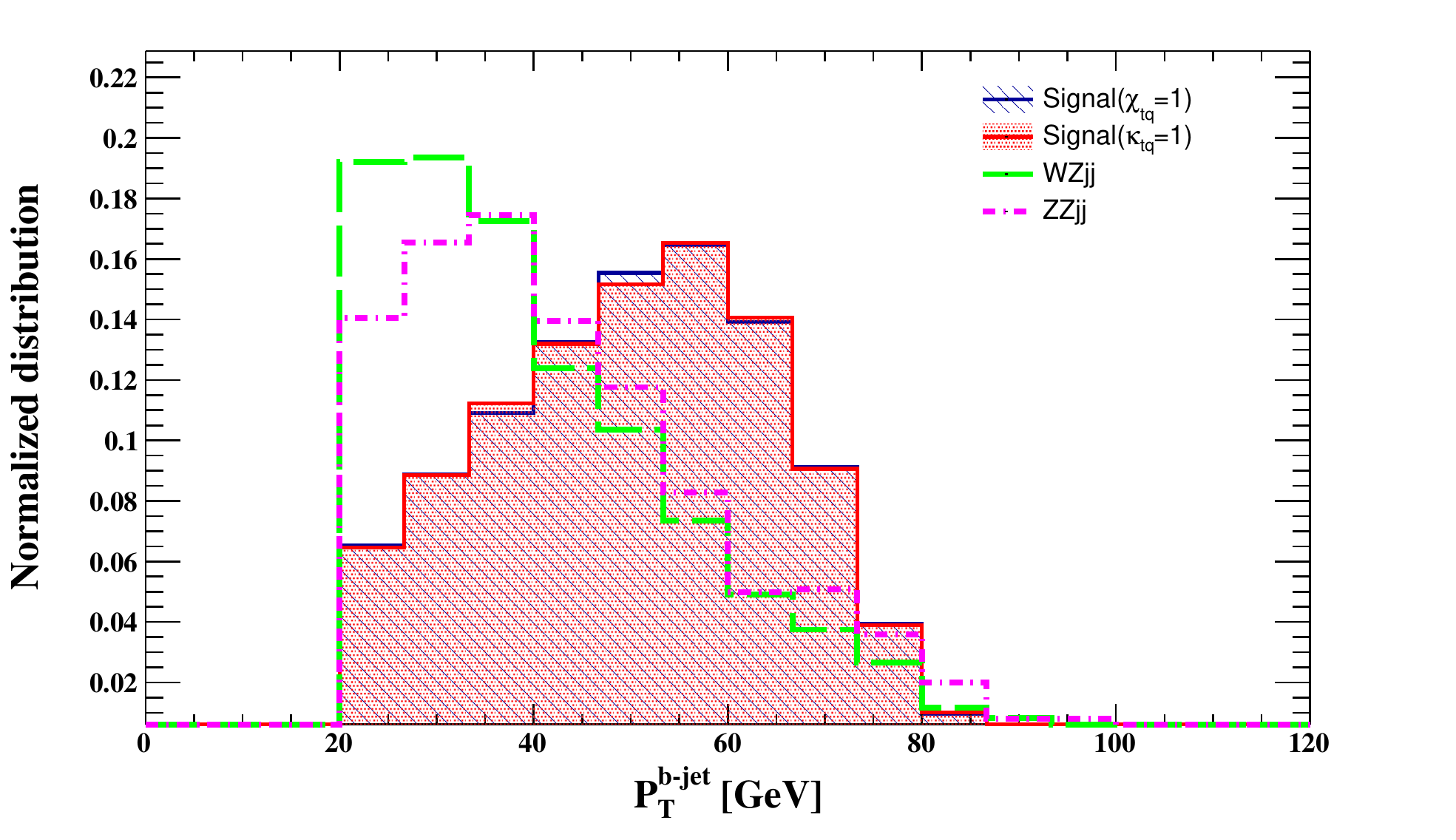}
		\includegraphics[width=0.45\textwidth]{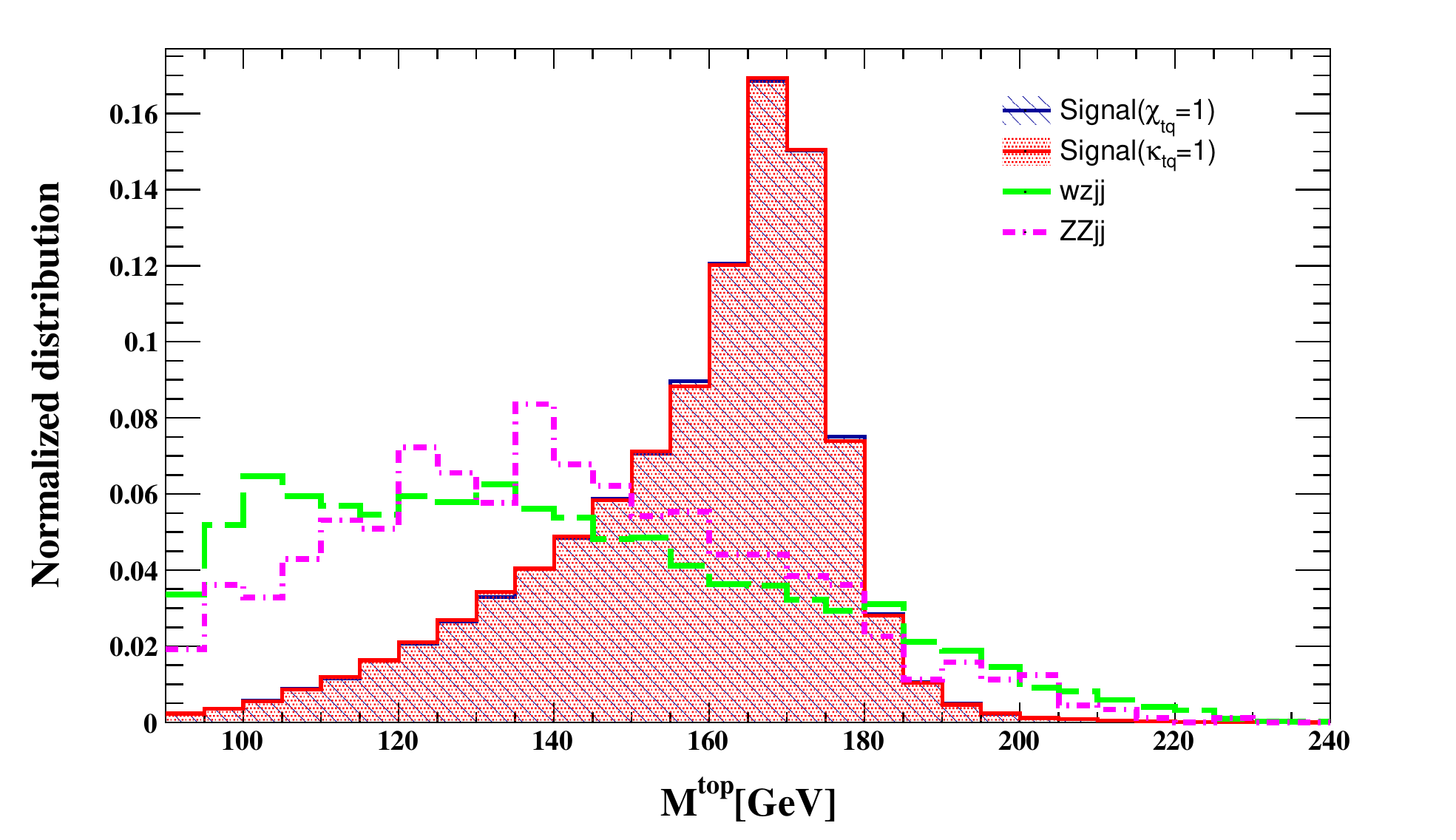}     \\
		\includegraphics[width=0.45\textwidth]{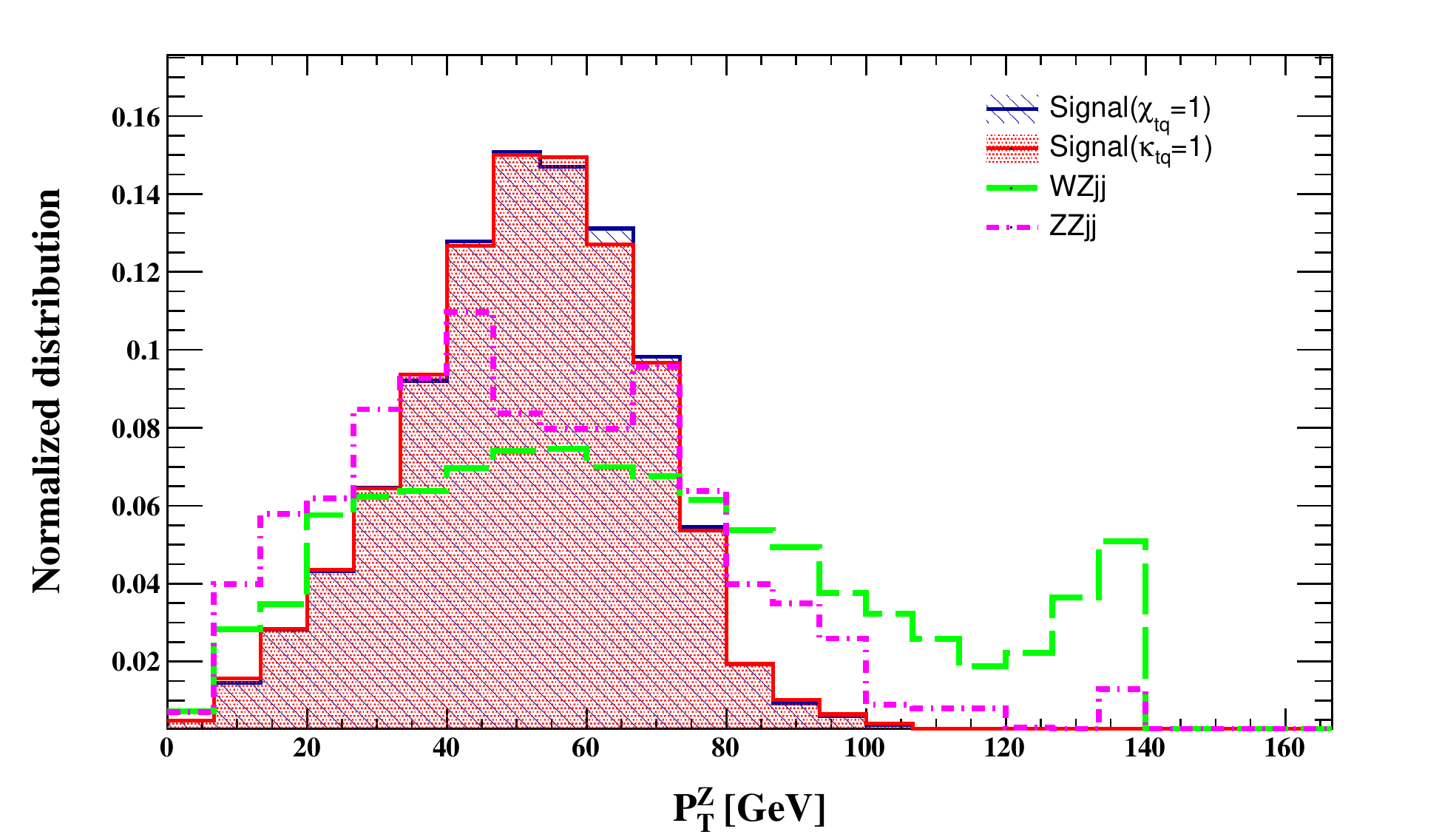}  
		\includegraphics[width=0.45\textwidth]{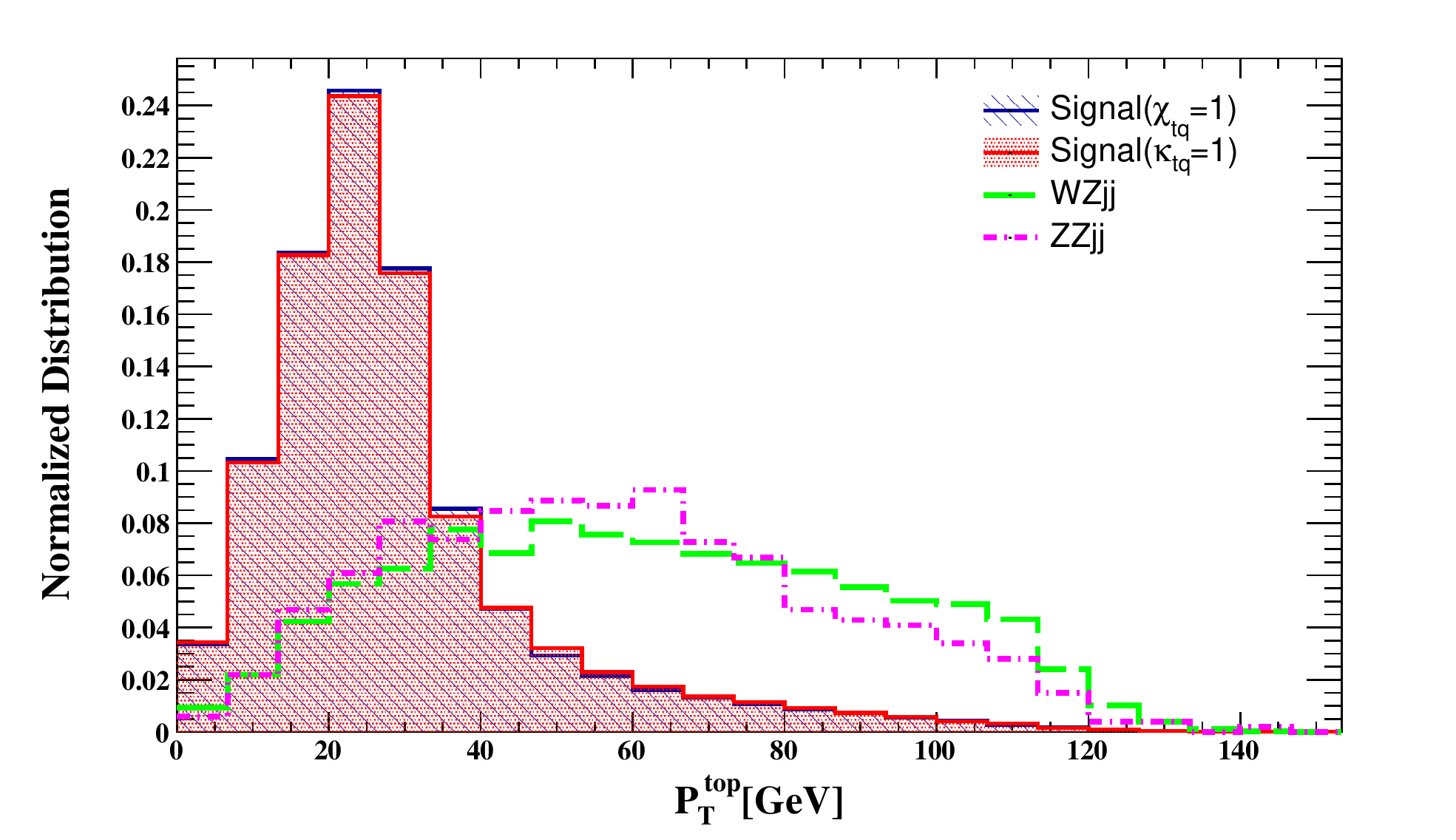}    \\
		\caption{The kinematic distributions of the top-Z-jet signal 
		(for $tqZ(\sigma_{\mu\nu})$ and $tqZ(\gamma_{\mu})$ scenarios) 
		and the main backgrounds, $WZjj$ and $ZZjj$, at the center of mass energy of $\sqrt {s} = 350$~GeV.}
		\label{Variables-Distributions-tjz}
	\end{center}
\end{figure*} 

Fig.~\ref{Variables-Distributions-tjz} illustrates some kinematic distributions of the top-Z-jet signal and 
the main background processes, $WZjj$ and $ZZjj$, 
after preliminary cuts and top quark and Z boson reconstruction.
Figures upper right and lower right show the mass and transverse momentum 
of the reconstructed top quark, respectively.
Figures upper left and lower left illustrate  the transverse momentum 
of the b-jet and the reconstructed Z boson, respectively.

To separate more background from signal events,  
we apply further cuts according to the distributions in Fig.~\ref{Variables-Distributions-tjz}. 
From the top quark's mass distribution (upper right), it is clear that the signal has a prominent 
peak around the nominal top mass.
A window cut on the reconstructed top quark mass ($140 < \rm{M}^{\rm{top}} <180$~GeV) 
causes to suppress backgrounds without a real top quark in the final state. 
According to the top quark's transverse momentum distribution (lower right), the signal's peak is around $20$~GeV, 
however, the main backgrounds have more spread distributions which extend up to around $120$~GeV.
Therefore, the transverse momentum of the top quark is required not to be greater than 30 GeV.
Moreover, according to the b-jet's transverse momentum distribution (upper left), 
we accept events with $p^{\rm{b-jet}}_{T} > 30$~GeV.
Also, we apply a cut on the Z boson's transverse momentum distribution (lower left)
to alleviate the number of background events which is required not to be greater than 80 GeV.

\begin{table}[t]
\begin{center}
\begin{footnotesize}
\begin{tabular}{|c|c|c|c|c|}
\hline
Cross-sections (in fb)     & \multicolumn{2}{|c|}{ Signal }        & \multicolumn{2}{|c|}{ Background }  \\ 
\hline
Cuts     &   $tqZ$ $(\sigma_{\mu\nu})$   &    $tqZ$ $(\gamma_{\mu})$    &  $WZjj$&$ZZjj$   \\
\hline  \hline
$p^{\rm{top}}_{T}<30~\rm{GeV}$ & $2.17(\kappa_{tq})^{2}$ &   $1.54(X_{tq})^{2}$     & $0.001$  &  $4.2\times 10^{-6}$  \\
$p^{Z}_{T}<80~\rm{GeV}$            & $2.14(\kappa_{tq})^{2}$ &   $1.52X_{tq})^{2}$     &   $0.0009$  &  $4\times 10^{-6}$   \\
$p^{\rm{b-jet}}_{T} > 30~\rm{GeV}$& $1.9(\kappa_{tq})^{2}$ &   $1.37(X_{tq})^{2}$   &  $0.0007$ &   $3.1\times 10^{-6}$   \\
$140< \rm{M}^{\rm{top}} <180~\rm{GeV}$& $1.7(\kappa_{tq})^{2}$  &   $1.14(X_{tq})^{2}$ &  $0.0003$  &   $1.7\times 10^{-6}$    \\ 
\hline
\end{tabular}
\end{footnotesize}
\caption{Cross sections for the  top-Z-jet signal (for $tqZ(\sigma_{\mu\nu})$ and $tqZ(\gamma_{\mu})$ scenarios) and the SM main background processes ($WZjj$ and $ZZjj$) after passing sequential secondary cuts at the center of mass energy of $\sqrt {s} = 350$~GeV.}
\label{secondCuts-tqZ}
\end{center}
\end{table}

The cross sections for the signal scenarios and the corresponding SM backgrounds 
after passing sequential secondary cuts at $\sqrt {s} = 350$~GeV are shown in Table~\ref{secondCuts-tqZ}.
The secondary cuts efficiencies for $tqZ$ $(\sigma_{\mu\nu})$ and $tqZ$ $(\gamma_{\mu})$  signal scenarios 
are $0.13$ and $0.12$ while those for $WZjj$ and $ZZjj$ backgrounds are $0.001$ and $0.0002$, respectively. 

The same method is used to find the secondary cuts and final efficiencies for $tq\gamma$ signal scenario.
Form the same kinematic distribution as before, we extract following cuts: 
$p^{\rm{top}}_{T}<80~\rm{GeV}$, $p^{Z}_{T}<80~\rm{GeV}$, $p^{\rm{b-jet}}_{T} > 30~\rm{GeV}$, 
and $140< \rm{M}^{\rm{top}} <180~\rm{GeV}$.Then, we find the corresponding efficiencies as; 
$\epsilon_{tq\gamma}=0.084$, $\epsilon_{WZjj}=0.002$, and $\epsilon_{ZZjj}=0.0004$.

\subsection{Sensitivity Estimation}
\label{sec:sensitivity}

In this subsection, we present the upper limits on the FCNC couplings ($tqZ$ and $tq\gamma$) 
at a confidence level (CL) of $95 \%$ by analyzing the single top quark production in association with a Z boson.
To estimate the sensitivity of the future electron-positron collider to these anomalous couplings,
we employ a Bayesian approach using a flat prior distribution~\cite{Tanabashi:2018oca,Backovic:2013bga}.
A Poisson distribution is assumed for the probability of observing $n_{\mathrm{obs}}$ events, 
\begin{equation}
L\left(n_{\mathrm{obs}}, n_{\mathrm{S}}, n_{\mathrm{B}} \right)=\frac{\left(n_{\mathrm{S}}+n_{\mathrm{B}}\right)^{n_{\mathrm{obs}}}}{n_{\mathrm{obs}} !} e^{-\left(n_{\mathrm{S}}+n_{\mathrm{B}}\right)}.
\end{equation}
Notice that in the above equation, $n_{\mathrm{S}}$ is defined by 
$n_{\mathrm{S}} = \epsilon_{\mathrm{S}} \times \cL \times \sigma_{\mathrm{S}}$, 
that $\epsilon_{\mathrm{S}}$ is the efficiency of the signal after all cuts 
and $\cL$ is the integrated luminosity.
For a given integrated luminosity and backgrounds cross section
(Table \ref{firstCuts-tqz}, first row), the number of expected background events are given 
by $n_{\mathrm{B}} = \epsilon_{\mathrm{B}} \times \cL \times \sigma_{\mathrm{B}}$, 
which $\epsilon_{\mathrm{B}}$ is the efficiency of the background processes 
after all the preliminary and secondary cuts.
The upper limit on the number of signal process of $e^-  e^+ \rightarrow t Z j$
can be obtained from the following equation at $95\%$ CL:
\begin{equation}
\frac{95}{100}=\frac{\int_{0}^{n_{\text {limit }}}L\left(n_{\mathrm{obs}}, n_{\mathrm{S}}, n_{\mathrm{B}} \right) \mathrm{d} n_{\mathrm{S}}}{\int_{0}^{\infty} L\left(n_{\mathrm{obs}}, n_{\mathrm{S}}, n_{\mathrm{B}} \right) \mathrm{d} n_{\mathrm{S}}}.
\end{equation}

We assume that the observed number of events at the detectors are consistent with the expected number of 
background events and solve above equation to find the upper limit on the number of signal events.
This limit can be translated to the upper limit on the signal cross section and then to the upper limits on the FCNC couplings.
\begin{table}[t]
\begin{footnotesize}
\begin{center}
\begin{tabular}{|c||c||c|}
\hline  Integrated luminosity & Coupling & Branching ratio \\
\hline 
\hline 
& $\kappa_{tq} < 0.07$ &$\rm{Br}(t \rightarrow q Z)\left(\sigma_{\mu \nu}\right)<1.8\times10^{-3}$ \\
\(300~\mathrm{fb}^{-1}\) & $X_{tq} < 0.09$ &$\rm{Br}(t \rightarrow q Z)\left(\gamma_{\mu}\right)<3.8\times10^{-3}$\\
& $\lambda_{tq} < 0.91$ &$\rm{Br}(t \rightarrow q \gamma)<3.5\times10^{-1}$ \\
\hline 
\hline 
& $\kappa_{tq} < 0.04$ &$\rm{Br}(t \rightarrow q Z)\left(\sigma_{\mu \nu}\right)<5.8\times10^{-4}$ \\
\(1~\mathrm{ab}^{-1}\) & $X_{tq} < 0.05$ &$\rm{Br}(t \rightarrow q Z)\left(\gamma_{\mu}\right)<1.2\times10^{-3}$\\
& $\lambda_{tq} < 0.50$ &$\rm{Br}(t \rightarrow q \gamma)<1.1\times10^{-1}$ \\
\hline 
& $\kappa_{tq} < 0.028$ &$\rm{Br}(t \rightarrow q Z)\left(\sigma_{\mu \nu}\right)<2.8\times10^{-4}$ \\
\(3~\mathrm{ab}^{-1}\) & $X_{tq} < 0.03$ &$\rm{Br}(t \rightarrow q Z)\left(\gamma_{\mu}\right)<4.2\times10^{-4}$\\
& $\lambda_{tq} < 0.33$ &$\rm{Br}(t \rightarrow q \gamma)<4.6\times10^{-2}$ \\
\hline 
\end{tabular}
\caption{The upper limits and the corresponding branching ratios at 95$\%$ CL 
on the $tqZ$ ($\sigma_{\mu \nu}$ and $\gamma_{\mu}$) and $tq\gamma$ 
at  the center of mass energies $\sqrt {s} = 350$~GeV for different integrated luminosity 
of $300~\rm{fb}^{-1}$, $1~\rm{ab}^{-1}$ and $3~\rm{ab}^{-1}$ by studying the top-Z-jet signal.}
\label{tab:LimitCouplings-tjZ}
\end{center}
\end{footnotesize}
\end{table}
Table \ref{tab:LimitCouplings-tjZ} presents the upper limits at 95$\%$ CL 
on the $tqZ$ ($\sigma_{\mu \nu}$ and $\gamma_{\mu}$) and $tq\gamma$ 
couplings at the center of mass energies $350$~GeV for different integrated luminosity 
of $300~\rm{fb}^{-1}$, $1~\rm{ab}^{-1}$ and $3~\rm{ab}^{-1}$ by studying the top-Z-jet channel.
As we expected before, the upper limit on the $\lambda_{tq}$ is looser than other anomalous couplings.
Then, these limits are translated to the corresponding branching which are shown in the 
table~\ref{tab:LimitCouplings-tjZ} for different integrated luminosities of data.
It is worth mentioning that these upper limits on the anomalous coupling can be translated to the bounds on 
$\Lambda/\sqrt{C_{i}}$ by using Eq.~\ref{Eq:Couplings}. 
According to our upper limits on the anomalous coupling, we can roughly say $\Lambda/\sqrt{C_{i}}$ 
should be greater than $\sim1.5$ TeV.

\section{Analysis Strategy for top-$\gamma$-jet signal}
\label{sec:tAjet}
In this section, we analyze the single top quark production in association with a photon process to search for 
the $tq\gamma$ and $tqZ$ FCNC couplings at the future lepton collider.
First of all, we describe event generation and simulation for the signal and background processes.
Then, we look at some kinematic distribution to separate the background events from the signal events. 
Finally, the upper limits which can be obtained from the top-$\gamma$-jet signal are given.
\subsection{Event Generation and Simulation}
\label{sec:EG}
Here, we define the top-$\gamma$-jet signal and related SM background processes 
and describe the event generation and simulation method.
In this analysis, we focus on the electron-positron collisions at the center of mass energy of 
$\sqrt {s} = 350$~GeV. 
The signal is defined as a single top production in association with 
a photon and a light jet ($e^-~e^+ \rightarrow t/\bar{t}~\gamma~j$), 
that top quark decays leptonically. 
Consequently, the final state includes a charged leptons (electron or muon), 
a hard photon, a b-jet, a light jet, and the missing energy due to the neutrino. 
Since the photon in the final state is a hard photon, we can handle it to suppress background contributions.
Some related Feynman diagrams  for top-$\gamma$-jet signal are demonstrated in Fig.~\ref{Feynman} 
($d$, $e$ and $f$).
Since our desirable signal (top-$\gamma$-jet) is raised up by all the $tqZ$ and $tq\gamma$ FCNC interactions,
again we employ three different scenarios to generate signal events; 
1) $\lambda_{tq}^{R}=\lambda_{tq}^{L}=\lambda_{tq} \neq 0$, 
2) $\kappa_{tq}^{R}=\kappa_{tq}^{L}=\kappa_{tq} \neq 0$,  and 3) $X_{tq}^{R}=X_{tq}^{L}=X_{tq} \neq 0$.
\begin{table}[t]
\begin{center}
\begin{footnotesize}
\begin{tabular}{|c|ccc|ccc|}
\hline
Cross-sections (in fb)     &  \multicolumn{3}{|c|}{ Signal }     &  \multicolumn{3}{|c|}{ Background }    \\ 
\hline
Cuts    & $tq\gamma $  &   $tqZ(\sigma_{\mu\nu})$   &    $tqZ(\gamma_{\mu})$    &  $Wjj\gamma$&$Zjj\gamma$& $t\bar{t}\gamma$   \\
\hline  \hline
Preselection cuts   &  $182(\lambda_{tq})^{2}$    & $230.6(\kappa_{tq})^{2}$ &   $29.3(X_{tq})^{2}$    & $84$  &  $2.9$  &   $0.0057$  \\
$1~\rm{light~jet} + 1\rm{b~jet}$    &  $98.16(\lambda_{tq})^{2}$   & $125.9(\kappa_{tq})^{2}$ &   $15.57(X_{tq})^{2}$     & $4.9$  &  $0.46$   &   $0.0045$   \\
$1~\rm{Photon}$   & $47.83(\lambda_{tq})^{2}$  & $33.3(\kappa_{tq})^{2}$ &   $4.5(X_{tq})^{2}$     &   $1.54$  &  $0.15$   &   $3 \times 10^{-5}$   \\
$1~\rm{lepton}$ & $38.26(\lambda_{tq})^{2}$  &  $26.34(\kappa_{tq})^{2}$ &   $3.6(X_{tq})^{2}$   &   $1.2$ &   $0.06$  &   $3\times 10^{-6}$    \\
$ \met>30~\rm{GeV}$  &   $28.4(\lambda_{tq})^{2}$    & $19.5(\kappa_{tq})^{2}$  &   $2.6X_{tq})^{2}$    &   $0.8$  &   $0.002$  &   $2 \times 10^{-6}$     \\ 
\hline
\end{tabular}
\end{footnotesize}
\caption{Cross sections for the  top-$\gamma$-jet signal and the SM background processes 
before and after passing sequential preliminary cuts at center of mass energy of $\sqrt {s} $ = 350 GeV.}
\label{firstCuts-tqA}
\end{center}
\end{table} 
 
According to the final state, we consider the following background processes;
$W^{\pm}jj\gamma$, $Zjj\gamma$, and $t\bar{t}\gamma$. 
The W-boson and Z-boson decay leptonically in all the  above backgrounds,
however, for $t\bar{t}\gamma$ semi-leptonic decay is assumed. 
In all cases, just decay to electron and muon are taken into account.
To generate and simulate the signal and background events, 
we follow the procedure which explained in subsection~\ref{sec:EG-tZjet}. 
 
In order to select the events, at least two jets with $p_{T} \geq 20$ GeV and $|\eta| < 2.4$ are required, 
which at least one of them should be tagged as a b-jet. 
Events including at least one isolated photon with $p_{T} \geq 30$ GeV and $|\eta| < 2.4$ are selected.
Furthermore, we require the events with exactly one charged leptons with $p_{T} \geq 10$ GeV and $|\eta| < 2.4$. 
Also, we choose events with missing transverse momentum greater than $30$ GeV. 

The cross sections including the branching ratios for three signal scenarios and 
the corresponding SM backgrounds after passing sequential preliminary cuts 
at $\sqrt {s} $ = 350 GeV are presented in Table \ref{firstCuts-tqA}.
The preliminary cuts efficiencies for $tq\gamma$, $tqZ(\sigma_{\mu\nu})$, and $tqZ(\gamma_{\mu})$ 
signal scenarios are $0.15$, $0.08$,  and $0.09$,  respectively.
The primary cuts efficiencies for $Wjj\gamma$, $Zjj\gamma$, and $t\bar{t}\gamma$ backgrounds
are at the order of $10^{-2}$, $10^{-4}$,  and $10^{-4}$, respectively. 
It is notable that these cuts can reduce significantly the number of background events. 

It is worth mentioning that there is another source for background for top-$\gamma$-jet channel.
Processes like  $W^{\pm}jj$, $Zjj$, and $t\bar{t}$ which don't have any real photon can mimic the signal
since their jets would be misidentified as photons in the detector.
The production cross sections for these background processes are large with respect to the 
other considered backgrounds, however, after above requirements, their efficiencies remarkably reduce.
As a result, here we neglect this source of background nevertheless in a more realistic detector simulation
their contribution should be taken into account.

\subsection{Separation of Signal over Background}
\label{sec:SBtAj}
In order to alleviate the number of the main SM background events, we follow the same way 
as described in subsection~\ref{sec:SBtZj}. Therefore, we focus on some kinematic distributions to find the
suitable secondary cuts.
Again at the first step, we need to reconstruct the top quark of the final state.
Since we have a clean final state, it is straightforward to reconstruct the top quark.
The top quark will be reconstructed by combining the charged lepton, the high energetic b-jet 
and the neutrino.
We show the kinematic distributions and describe our obtained secondary cuts and 
limits for $\lambda_{tq}$ anomalous coupling. Then, we will follow the same procedure 
to find the upper limits on the $\kappa_{tq}$ and $X_{tq}$ FCNC couplings.
\begin{figure*}[t]
	\begin{center}
		\includegraphics[width=0.45\textwidth]{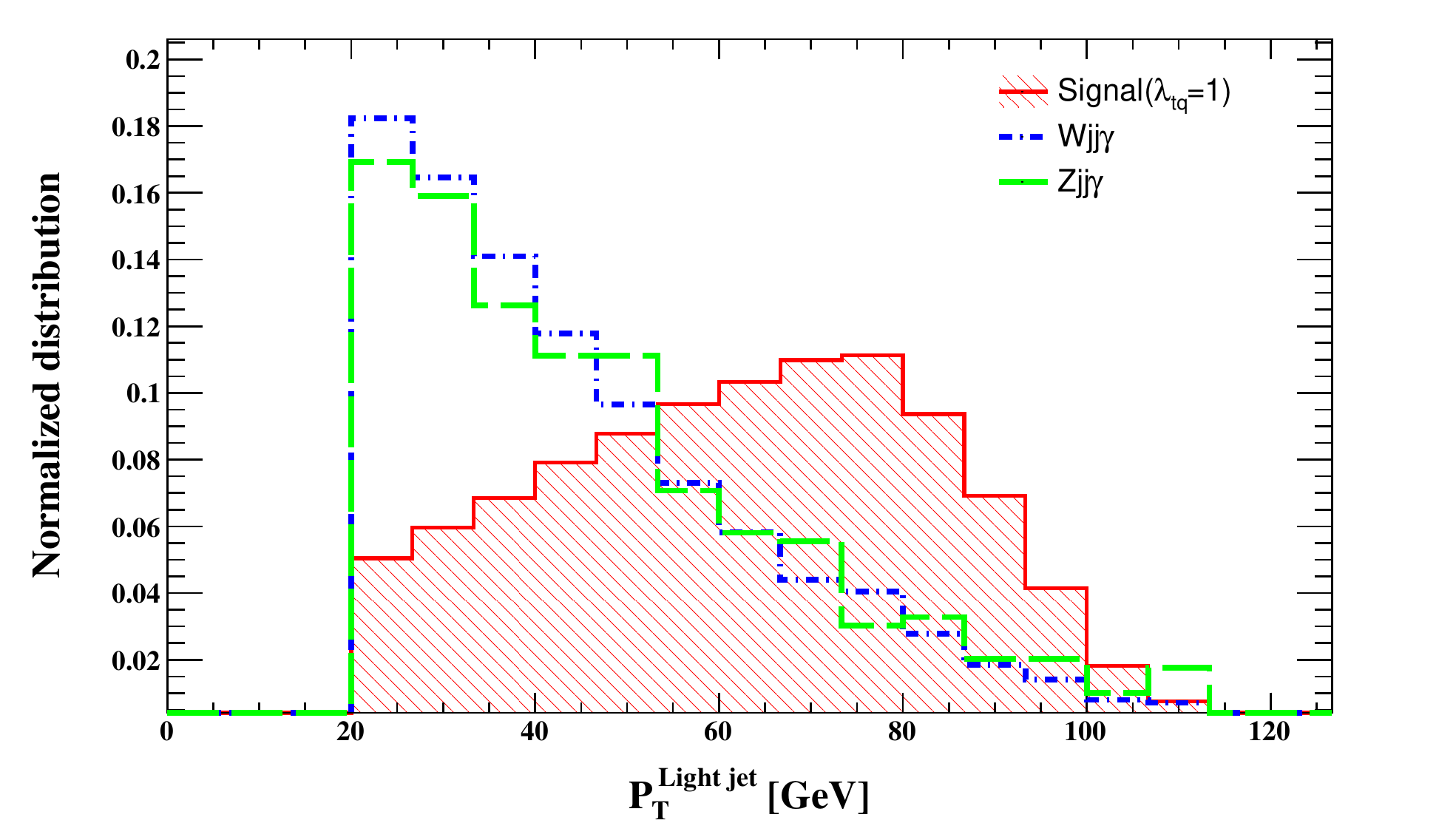}
		\includegraphics[width=0.45\textwidth]{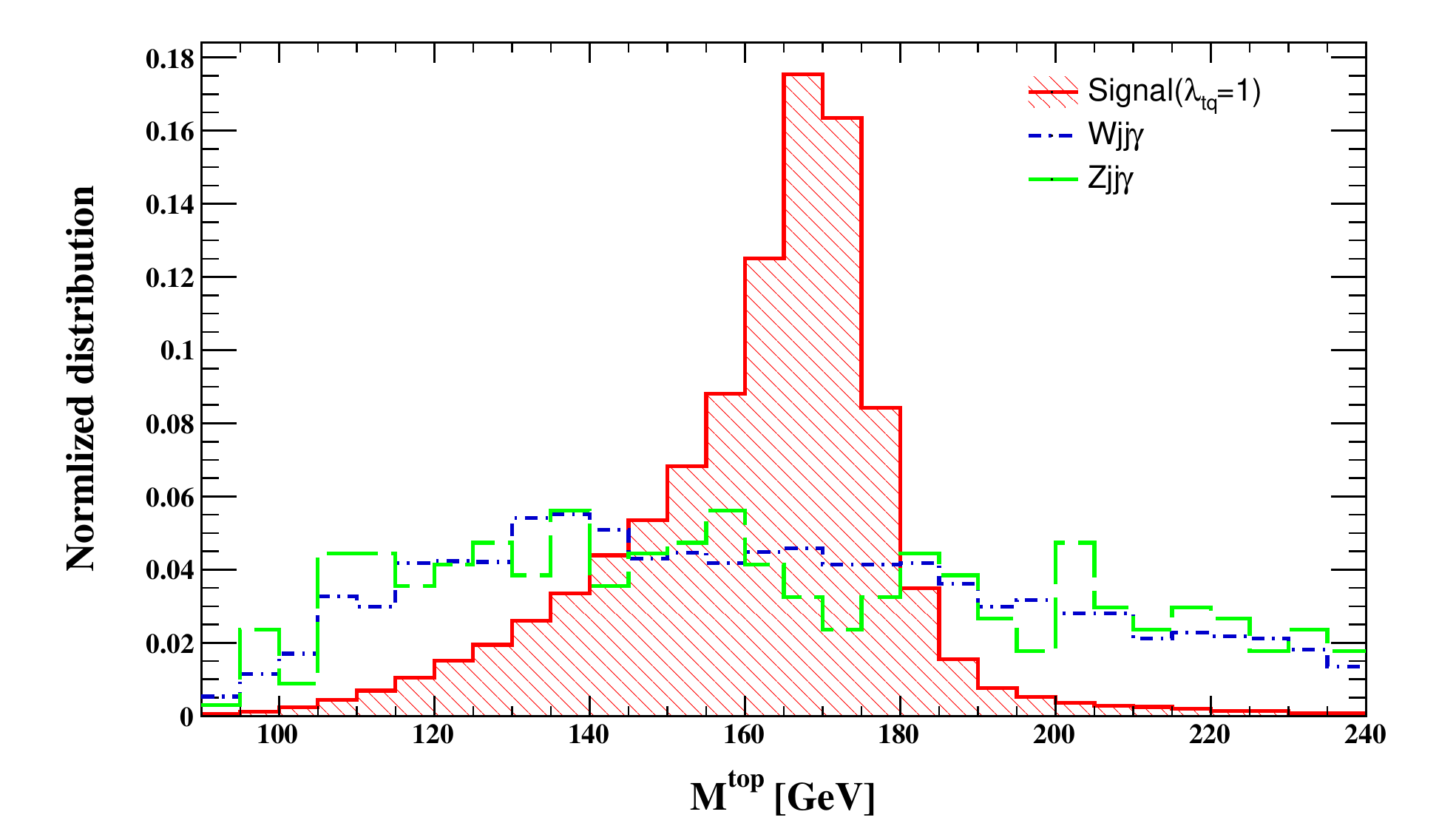}  \\
	     \includegraphics[width=0.45\textwidth]{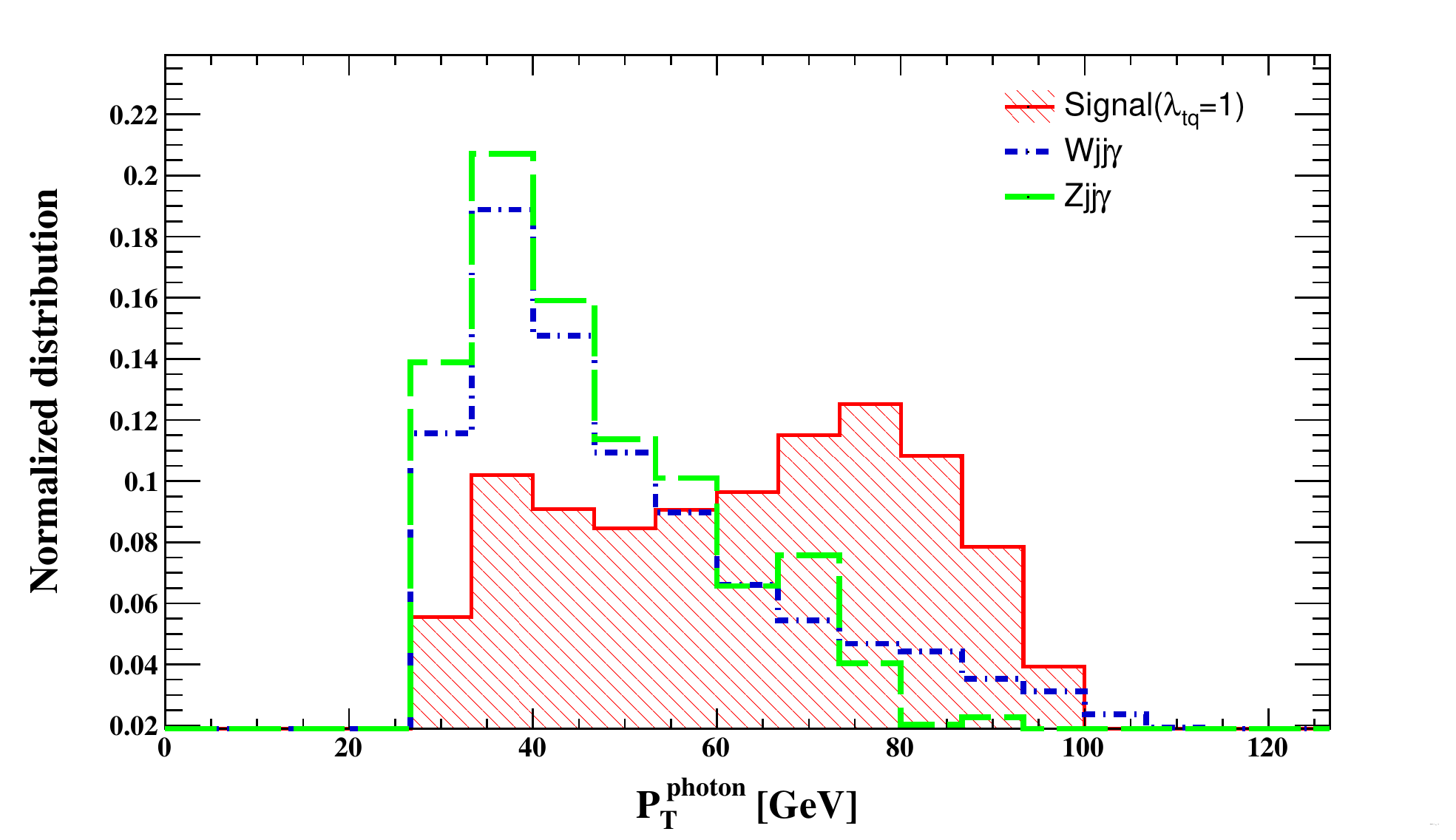}   
		\includegraphics[width=0.45\textwidth]{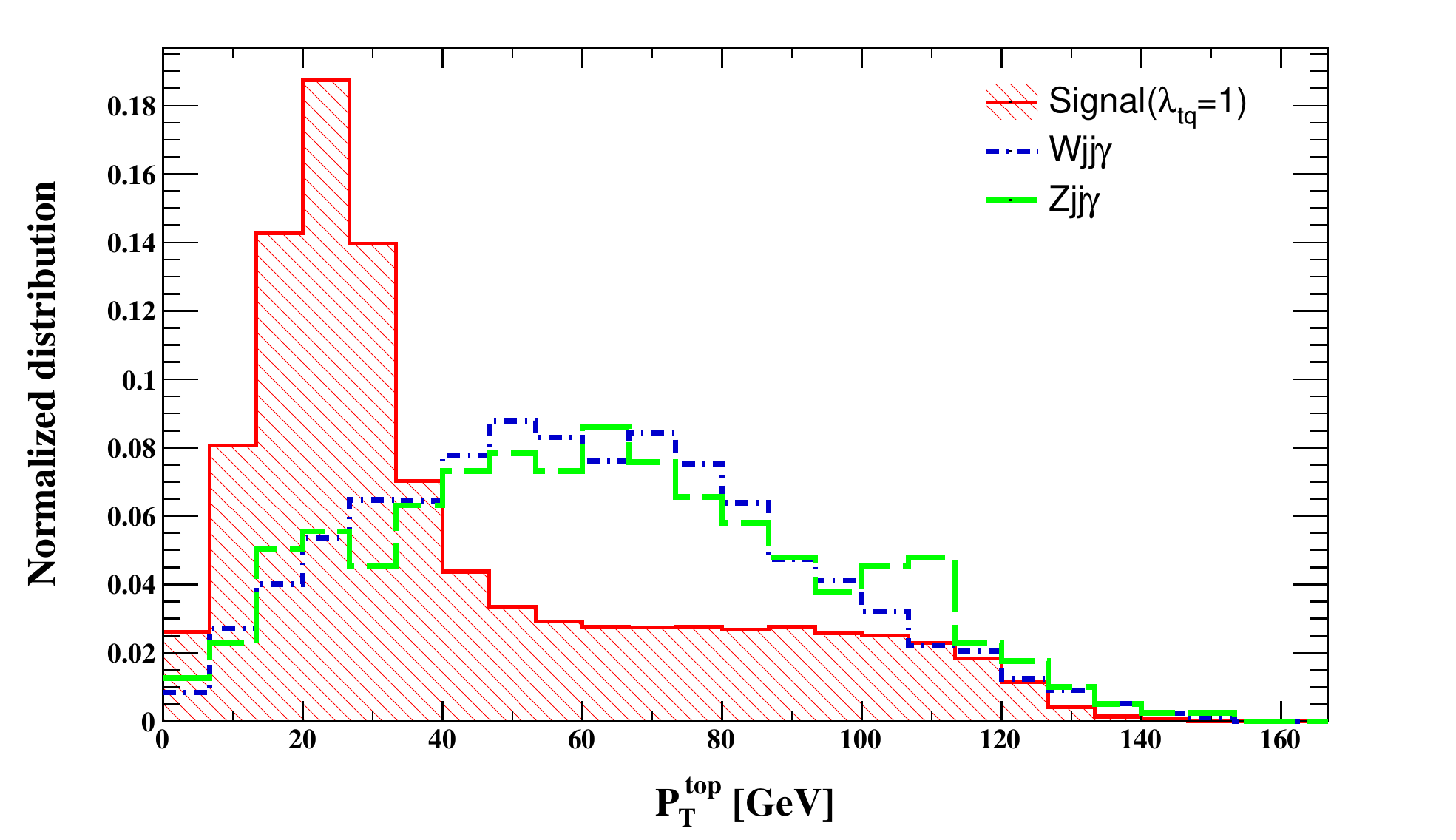}    \\
		\caption{The kinematic distributions of the top-$\gamma$-jet signal
		(for $tq\gamma$ scenario) 
		and the main backgrounds; $Wjj\gamma$ and $Zjj\gamma$
		 at the center of mass energy of $\sqrt {s} = 350$~GeV.}
		\label{Variables-Distributions-tjA}
	\end{center}
\end{figure*}

Fig.~\ref{Variables-Distributions-tjA} demonstrates different kinematic distributions of the top-$\gamma$-jet signal and 
some SM background processes ($Wjj \gamma$ and $Zjj\gamma$)
after preselection cuts and top quark reconstruction.
Figures upper right and lower right demonstrate mass and the transverse momentum 
of the reconstructed top quark, respectively.
Figures upper left and lower left illustrate the transverse momentum 
of the light-jet and the photon, respectively.

For suppressing more background events,  
we apply further cuts according to distributions in Fig.~\ref{Variables-Distributions-tjA}. 
The top quark mass distribution (upper right) for the signal shows a clear peak around $170$~GeV.
Therefore, we applied a window cut on the reconstructed top quark mass ($140 < \rm{M}^{\rm{top}} <180$~GeV). 
From the top quark transverse momentum distribution (lower right), the signal's peak is around $20$~GeV, 
however, the main backgrounds have more spread distributions which extended up to around $160$~GeV.
So, the transverse momentum of the top quark is required not to be greater than 40 GeV.

\begin{table}[t]
\begin{center}
\begin{footnotesize}
\begin{tabular}{|c|c|ccc|}
\hline
Cross-sections (in fb)     &  Signal        & \multicolumn{3}{|c|}{ Background }    \\ 
\hline
Cuts     &    $tq\gamma $  &  $Wjj\gamma$&$Zjj\gamma$& $t\bar{t}\gamma$  \\
\hline  \hline
$p^{\rm{top}}_{T}<40~\rm{GeV}$            &  $18.4(\lambda_{tq})^{2}$  & $0.2$  &  $5.7\times10^{-4}$  &   $1\times10^{-6}$  \\
$p^{\gamma}_{T}>40~\rm{GeV}$                  &  $17(\lambda_{tq})^{2}$ & $0.13$& $3.3\times10^{-4}$  &  $4.6\times10^{-7}$     \\
$p^{\rm{light-jet}}_{T} > 40~\rm{GeV}$        &  $15(\lambda_{tq})^{2}$  &  $0.07$ &   $1.8\times10^{-4}$  &   $2.6\times10^{-7}$   \\
$140 < \rm{M}^{\rm{top}} <180~\rm{GeV}$  &  $12.6(\lambda_{tq})^{2}$  &  $0.02$  &   $2.3\times10^{-5}$  &   $2\times10^{-7}$   \\ 
\hline
\end{tabular}
\end{footnotesize}
\caption{Cross sections for the top-$\gamma$-jet signal (for $tq\gamma$ scenario) 
and the SM background processes after passing sequential secondary 
cuts at the center of mass energy of $\sqrt {s} = 350$~GeV.}
\label{secondCuts-tqA}
\end{center}
\end{table}

Furthermore, according to the light-jet's transverse momentum distribution (upper left), 
we desire events with $p^{\rm{light-jet}}_{T} > 40$~GeV.
Also, we apply a cut on the hard photon transverse momentum distribution (lower left)
to reduce more number of background events which is required to be greater than 40 GeV.
In addition, according to the jet and b-jet multiplicities, we choose events with number of 
jets and b-jet less than 3 and 2, respectively.
This requirement can suppress a large number of background events with a high number of jets.

The cross sections for the $tq\gamma$ signal scenarios and the corresponding SM backgrounds
after passing sequential secondary cuts are shown in Table~\ref{secondCuts-tqA}.
The final efficiencies for the $tq\gamma$ signal scenario is $0.07$ while those 
for $Wjj\gamma$, $Zjj\gamma$, and $t\bar{t}\gamma$ backgrounds
are at the order of $10^{-4}$, $10^{-6}$, and $10^{-5}$, respectively. 
The same method is used to find the secondary cuts and final efficiencies for 
$tqZ(\sigma_{\mu\nu})$ and $tqZ(\gamma_{\mu})$ signal scenarios.
In the following, these final efficiencies are used to estimate the upper limits on the anomalous couplings.

\subsection{Sensitivity Estimation}
\label{sec:sensitivitytqA}
In this part, the obtained upper limits on the FCNC couplings ($tqZ$ and $tq\gamma$) 
at $95 \%$ CL by analyzing the single top quark production in association with photon are presented.
Again, we follow the same procedure which explained in subsection \ref{sec:sensitivity}. 
\begin{table}[t]
\begin{footnotesize}
\begin{center}
\begin{tabular}{|c||c||c|}
\hline  Integrated luminosity & Coupling & Branching ratio \\
\hline 
\hline 
& $\kappa_{tq} < 0.065$ &$\rm{Br}(t \rightarrow q Z)\left(\sigma_{\mu \nu}\right)<1.5\times10^{-3}$ \\
\(300~\mathrm{fb}^{-1}\) & $X_{tq} < 0.15$ &$\rm{Br}(t \rightarrow q Z)\left(\gamma_{\mu}\right)<1.1\times10^{-2}$\\
& $\lambda_{tq} < 0.04$ &$\rm{Br}(t \rightarrow q \gamma)<6.8\times10^{-4}$ \\
\hline 
\hline 
& $\kappa_{tq} < 0.047$ &$\rm{Br}(t \rightarrow q Z)\left(\sigma_{\mu \nu}\right)<8\times10^{-4}$ \\
\(1~\mathrm{ab}^{-1}\) & $X_{tq} < 0.11$ &$\rm{Br}(t \rightarrow q Z)\left(\gamma_{\mu}\right)<5.7\times10^{-3}$\\
& $\lambda_{tq} < 0.03$ &$\rm{Br}(t \rightarrow q \gamma)<3.87\times10^{-4}$ \\
\hline 
& $\kappa_{tq} < 0.035$ &$\rm{Br}(t \rightarrow q Z)\left(\sigma_{\mu \nu}\right)<4.5\times10^{-4}$ \\
\(3~\mathrm{ab}^{-1}\) & $X_{tq} < 0.085$ &$\rm{Br}(t \rightarrow q Z)\left(\gamma_{\mu}\right)<3.4\times10^{-3}$\\
& $\lambda_{tq} < 0.02$ &$\rm{Br}(t \rightarrow q \gamma)<1.7\times10^{-4}$ \\
\hline 
\end{tabular}
\caption{The upper limits and the corresponding branching ratios at 95$\%$ CL 
on the $tqZ$ ($\sigma_{\mu \nu}$ and $\gamma_{\mu}$) and $tq\gamma$ 
at  the center of mass energies $\sqrt {s} = 350$~GeV for different integrated luminosity 
of $300~\rm{fb}^{-1}$, $1~\rm{ab}^{-1}$ and $3~\rm{ab}^{-1}$ by studying the top-$\gamma$-jet signal.}
\label{tab:LimitCouplings-tjgamma}
\end{center}
\end{footnotesize}
\end{table}

Table \ref{tab:LimitCouplings-tjgamma} shows the upper limits and the corresponding branching ratios at 95$\%$ CL 
on the $tqZ$ ($\sigma_{\mu \nu}$ and $\gamma_{\mu}$) and $tq\gamma$ 
at the center of mass energy of 350 GeV for different integrated luminosities 
of  300 $\rm{fb}^{-1}$, 1 $\rm{ab}^{-1}$ and 3 $\rm{ab}^{-1}$ using top-$\gamma$-jet signal.
Also, we do the same study on top-$\gamma$-jet channel
at the center of mass energy of 240 GeV for the integrated luminosity 
of $10~\rm{ab}^{-1}$ and find upper limits at $95\%$ CL on the branching ratio 
as follows; $\rm{Br}(t \rightarrow q \gamma) < 2.2\times10^{-2}$.
It is worth mentioning that the cross section of $t\bar{t}$ at the threshold is too small by a factor of three.
So we consider the threshold resummation effect on the $t\bar{t}$ cross section, however, 
these effects change our result by less than 5\%.

\section{Summary}
\label{sec:summary}

Flavour Changing Neutral Current (FCNC) interactions in the top quark sector are highly 
suppressed in the Standard Model (SM)
framework. The predicted branching ratios for different top FCNC decay modes are of $\cO(10^{-14})$ in the SM,
while there are some Beyond SM models that anticipate large values for these branching ratios.
Therefore, any sign of top quark FCNC interaction would be a promising indication of the existence of new physics.
On the other hand, future electron-positron colliders with high luminosity
are able to produce large numbers of top quark and give us this opportunity to probe top quark properties with
high precision.

In this paper, we study the single top quark production in association with a neutral gauge boson (Z boson or photon)
to search for $tqZ$ and $tq\gamma$ anomalous couplings at the future electron-positron collider.
Since many BSMs predict the FCNC interactions with the same signatures at particle colliders in many cases, 
using the model independent approach is more useful. So, we employ 
the Standard Model Effective Field Theory (SMEFT) approach in this study.
We perform two separate analyses for top-Z-jet and  top-$\gamma$-jet channels.

In the top-Z-jet study, we consider leptonic decay of top quark and Z boson.
As a result, we have three charged leptons (electron or muon), a light-jet, a b-jet 
and missing energy in the final state.
Since all the FCNC couplings $tqZ(\sigma_{\mu \nu})$, $tqZ(\gamma_{\mu})$ and $tq\gamma$
can produce our favourable signal, we categorize our signal in three scenarios.
We consider all the main SM background processes for this channel.  
Also, the parton showering and hadronization are taken into account as well as fast detector simulation. 
Then, we use a cut-and-count technique to separate the signal from the background events.
The upper limits at $95\%$ CL are obtained on the anomalous couplings and 
the corresponding branching ratios for different the integrated luminosity at the center of mass energy of $350$ GeV.
For example, it is found at the $95\%$ CL with $3~\rm{ab}^{-1}$ luminosity of data; 
$\rm{Br}(t \rightarrow q Z)\left(\sigma_{\mu \nu}\right)<2.8\times10^{-4}$,
$\rm{Br}(t \rightarrow q Z)\left(\gamma_{\mu}\right)<4.2\times10^{-4}$,
and $\rm{Br}(t \rightarrow q \gamma)<4.6\times10^{-2}$.

In the top-$\gamma$-jet study, leptonically decay of top quark is assumed.
So, the final state consists of one charged leptons (electron or muon), 
a hard photon, a light-jet, a b-jet and the missing energy.
The upper limits at $95\%$ CL are obtained for $3~\rm{ab}^{-1}$ luminosity of data
at the center of mass energy $350$ GeV are as follows;
$\rm{Br}(t \rightarrow q Z)\left(\sigma_{\mu \nu}\right)<4.5\times10^{-4}$,
$\rm{Br}(t \rightarrow q Z)\left(\gamma_{\mu}\right)<3.4\times10^{-3}$,
and $\rm{Br}(t \rightarrow q \gamma)<1.7\times10^{-4}$.
Also, it is worth mentioning here that by using polarized beams 
the obtained upper limits can be improved~\cite{MoortgatPick:2005cw}.
Our obtained results would be improved significantly by using the statistical  techniques for separating more signal 
from background events such as multivariate analysis.
The combination of the obtained upper limits from the top quark production associated with a 
neutral gauge boson with other channels also can lead to stringent bounds on the 
FCNC couplings($tqZ$ ana $tq\gamma$), particularly tighter limits would be expected 
from combination with limits coming from the top+jet final state.

\section*{Acknowledgments}
\label{sec:ack}
Authors would like to especially thank Mojtaba Mohammadi Najafabadi and Hamzeh Khanpour for many useful discussions and their invaluable comments on the draft. Authors are also thankful School of Particles and Accelerators, Institute for Research in Fundamental Sciences (IPM) for hospitality.

\bibliographystyle{JHEP}
\bibliography{FCNCRef}

\providecommand{\href}[2]{#2}\begingroup\raggedright\begin{thebibliography}{10}

\bibitem{Glashow:1970gm}
S.~L. Glashow, J.~Iliopoulos, and L.~Maiani, {\it {Weak Interactions with
  Lepton-Hadron Symmetry}},  {\em Phys. Rev.} {\bf D2} (1970) 1285--1292.

\bibitem{Agashe:2013hma}
{\bf Top Quark Working Group} Collaboration, K.~Agashe {\em et.~al.}, {\it
  {Working Group Report: Top Quark}},  in {\em {Proceedings, 2013 Community
  Summer Study on the Future of U.S. Particle Physics: Snowmass on the
  Mississippi (CSS2013): Minneapolis, MN, USA, July 29-August 6, 2013}}, 2013.
\newblock \href{http://arxiv.org/abs/1311.2028}{{\tt 1311.2028}}.

\bibitem{Luke:1993cy}
M.~E. Luke and M.~J. Savage, {\it {Flavor changing neutral currents in the
  Higgs sector and rare top decays}},  {\em Phys. Lett.} {\bf B307} (1993)
  387--393, [\href{http://arxiv.org/abs/hep-ph/9303249}{{\tt hep-ph/9303249}}].

\bibitem{Atwood:1996vj}
D.~Atwood, L.~Reina, and A.~Soni, {\it {Phenomenology of two Higgs doublet
  models with flavor changing neutral currents}},  {\em Phys. Rev.} {\bf D55}
  (1997) 3156--3176, [\href{http://arxiv.org/abs/hep-ph/9609279}{{\tt
  hep-ph/9609279}}].

\bibitem{Liu:2004qw}
J.~J. Liu, C.~S. Li, L.~L. Yang, and L.~G. Jin, {\it {$t \to c$V via SUSY FCNC
  couplings in the unconstrained MSSM}},  {\em Phys. Lett.} {\bf B599} (2004)
  92--101, [\href{http://arxiv.org/abs/hep-ph/0406155}{{\tt hep-ph/0406155}}].

\bibitem{Delepine:2004hr}
D.~Delepine and S.~Khalil, {\it {Top flavor violating decays in general
  supersymmetric models}},  {\em Phys. Lett.} {\bf B599} (2004) 62--74,
  [\href{http://arxiv.org/abs/hep-ph/0406264}{{\tt hep-ph/0406264}}].

\bibitem{Li:1993mg}
C.~S. Li, R.~J. Oakes, and J.~M. Yang, {\it {Rare decay of the top quark in the
  minimal supersymmetric model}},  {\em Phys. Rev.} {\bf D49} (1994) 293--298.
  [Erratum: Phys. Rev.D56,3156(1997)].

\bibitem{Lopez:1997xv}
J.~L. Lopez, D.~V. Nanopoulos, and R.~Rangarajan, {\it {New supersymmetric
  contributions to $t \to c$ V}},  {\em Phys. Rev.} {\bf D56} (1997)
  3100--3106, [\href{http://arxiv.org/abs/hep-ph/9702350}{{\tt
  hep-ph/9702350}}].

\bibitem{Yang:1997dk}
J.~M. Yang, B.-L. Young, and X.~Zhang, {\it {Flavor changing top quark decays
  in r parity violating SUSY}},  {\em Phys. Rev.} {\bf D58} (1998) 055001,
  [\href{http://arxiv.org/abs/hep-ph/9705341}{{\tt hep-ph/9705341}}].

\bibitem{AguilarSaavedra:2004wm}
J.~A. Aguilar-Saavedra, {\it {Top flavor-changing neutral interactions:
  Theoretical expectations and experimental detection}},  {\em Acta Phys.
  Polon.} {\bf B35} (2004) 2695--2710,
  [\href{http://arxiv.org/abs/hep-ph/0409342}{{\tt hep-ph/0409342}}].

\bibitem{AguilarSaavedra:2000aj}
J.~A. Aguilar-Saavedra and G.~C. Branco, {\it {Probing top flavor changing
  neutral scalar couplings at the CERN LHC}},  {\em Phys. Lett.} {\bf B495}
  (2000) 347--356, [\href{http://arxiv.org/abs/hep-ph/0004190}{{\tt
  hep-ph/0004190}}].

\bibitem{AguilarSaavedra:2002ns}
J.~A. Aguilar-Saavedra and B.~M. Nobre, {\it {Rare top decays $t \to c$ gamma,
  $t \to c g$ and CKM unitarity}},  {\em Phys. Lett.} {\bf B553} (2003)
  251--260, [\href{http://arxiv.org/abs/hep-ph/0210360}{{\tt hep-ph/0210360}}].

\bibitem{Khatibi:2015aal}
S.~Khatibi and M.~Mohammadi~Najafabadi, {\it {Constraints on top quark flavor
  changing neutral currents using diphoton events at the LHC}},  {\em Nucl.
  Phys.} {\bf B909} (2016) 607--618,
  [\href{http://arxiv.org/abs/1511.00220}{{\tt 1511.00220}}].

\bibitem{Khanpour:2014xla}
H.~Khanpour, S.~Khatibi, M.~Khatiri~Yanehsari, and M.~Mohammadi~Najafabadi,
  {\it {Single top quark production as a probe of anomalous $tq\gamma$ and
  $tqZ$ couplings at the FCC-ee}},  {\em Phys. Lett.} {\bf B775} (2017) 25--31,
  [\href{http://arxiv.org/abs/1408.2090}{{\tt 1408.2090}}].

\bibitem{Khatibi:2014via}
S.~Khatibi and M.~Mohammadi~Najafabadi, {\it {Probing the Anomalous FCNC
  Interactions in Top-Higgs Final State and Charge Ratio Approach}},  {\em
  Phys. Rev.} {\bf D89} (2014), no.~5 054011,
  [\href{http://arxiv.org/abs/1402.3073}{{\tt 1402.3073}}].

\bibitem{Gao:2011fx}
J.~Gao, C.~S. Li, L.~L. Yang, and H.~Zhang, {\it {Search for anomalous top
  quark production at the early LHC}},  {\em Phys. Rev. Lett.} {\bf 107} (2011)
  092002, [\href{http://arxiv.org/abs/1104.4945}{{\tt 1104.4945}}].

\bibitem{Khanpour:2019qnw}
H.~Khanpour, {\it {Probing top quark FCNC couplings in the triple-top signal at
  the high energy LHC and future circular collider}},  {\em Nucl. Phys. B} {\bf
  958} (2020) 115141, [\href{http://arxiv.org/abs/1909.03998}{{\tt
  1909.03998}}].

\bibitem{Hesari:2015oya}
H.~Hesari, H.~Khanpour, and M.~Mohammadi~Najafabadi, {\it {Direct and Indirect
  Searches for Top-Higgs FCNC Couplings}},  {\em Phys. Rev.} {\bf D92} (2015),
  no.~11 113012, [\href{http://arxiv.org/abs/1508.07579}{{\tt 1508.07579}}].

\bibitem{Hesari:2014eua}
H.~Hesari, H.~Khanpour, M.~Khatiri~Yanehsari, and M.~Mohammadi~Najafabadi, {\it
  {Probing the Top Quark Flavour-Changing Neutral Current at a Future
  Electron-Positron Collider}},  {\em Adv. High Energy Phys.} {\bf 2014} (2014)
  476490, [\href{http://arxiv.org/abs/1412.8572}{{\tt 1412.8572}}].

\bibitem{Etesami:2010ig}
S.~M. Etesami and M.~Mohammadi~Najafabadi, {\it {Study of anomalous top quark
  FCNC interactions via $tW$-channel of single top}},  {\em Phys. Rev.} {\bf
  D81} (2010) 117502, [\href{http://arxiv.org/abs/1006.1717}{{\tt 1006.1717}}].

\bibitem{Aad:2015uza}
{\bf ATLAS} Collaboration, G.~Aad {\em et.~al.}, {\it {Search for
  flavour-changing neutral current top-quark decays to $qZ$ in $pp$ collision
  data collected with the ATLAS detector at $\sqrt s =8$ TeV}},  {\em Eur.
  Phys. J.} {\bf C76} (2016), no.~1 12,
  [\href{http://arxiv.org/abs/1508.05796}{{\tt 1508.05796}}].

\bibitem{Aad:2015gea}
{\bf ATLAS} Collaboration, G.~Aad {\em et.~al.}, {\it {Search for single
  top-quark production via flavour-changing neutral currents at 8 TeV with the
  ATLAS detector}},  {\em Eur. Phys. J.} {\bf C76} (2016), no.~2 55,
  [\href{http://arxiv.org/abs/1509.00294}{{\tt 1509.00294}}].

\bibitem{Aad:2012ij}
{\bf ATLAS} Collaboration, G.~Aad {\em et.~al.}, {\it {A search for flavour
  changing neutral currents in top-quark decays in $pp$ collision data
  collected with the ATLAS detector at $\sqrt{s}=7$ TeV}},  {\em JHEP} {\bf 09}
  (2012) 139, [\href{http://arxiv.org/abs/1206.0257}{{\tt 1206.0257}}].

\bibitem{Khachatryan:2015att}
{\bf CMS} Collaboration, V.~Khachatryan {\em et.~al.}, {\it {Search for
  Anomalous Single Top Quark Production in Association with a Photon in $pp$
  Collisions at $ \sqrt{s}=8 $ TeV}},  {\em JHEP} {\bf 04} (2016) 035,
  [\href{http://arxiv.org/abs/1511.03951}{{\tt 1511.03951}}].

\bibitem{Aad:2019pxo}
{\bf ATLAS} Collaboration, G.~Aad {\em et.~al.}, {\it {Search for
  flavour-changing neutral currents in processes with one top quark and a
  photon using 81 fb$^{-1}$ of $pp$ collisions at $\sqrt{s} = 13$ TeV with the
  ATLAS experiment}},  {\em Phys. Lett.} {\bf B800} (2020) 135082,
  [\href{http://arxiv.org/abs/1908.08461}{{\tt 1908.08461}}].

\bibitem{Aaboud:2018nyl}
{\bf ATLAS} Collaboration, M.~Aaboud {\em et.~al.}, {\it {Search for
  flavour-changing neutral current top-quark decays $t\to qZ$ in proton-proton
  collisions at $\sqrt{s}=13$ TeV with the ATLAS detector}},  {\em JHEP} {\bf
  07} (2018) 176, [\href{http://arxiv.org/abs/1803.09923}{{\tt 1803.09923}}].

\bibitem{Chatrchyan:2012hqa}
{\bf CMS} Collaboration, S.~Chatrchyan {\em et.~al.}, {\it {Search for Flavor
  Changing Neutral Currents in Top Quark Decays in pp Collisions at 7 TeV}},
  {\em Phys. Lett.} {\bf B718} (2013) 1252--1272,
  [\href{http://arxiv.org/abs/1208.0957}{{\tt 1208.0957}}].

\bibitem{Chatrchyan:2013nwa}
{\bf CMS} Collaboration, S.~Chatrchyan {\em et.~al.}, {\it {Search for
  Flavor-Changing Neutral Currents in Top-Quark Decays $t \to Zq$ in $pp$
  Collisions at $\sqrt{s}=8$ TeV}},  {\em Phys. Rev. Lett.} {\bf 112} (2014),
  no.~17 171802, [\href{http://arxiv.org/abs/1312.4194}{{\tt 1312.4194}}].

\bibitem{Abazov:2011qf}
{\bf D0} Collaboration, V.~M. Abazov {\em et.~al.}, {\it {Search for Flavor
  Changing Neutral Currents in Decays of Top Quarks}},  {\em Phys. Lett.} {\bf
  B701} (2011) 313--320, [\href{http://arxiv.org/abs/1103.4574}{{\tt
  1103.4574}}].

\bibitem{Abramowicz:2011tv}
{\bf ZEUS} Collaboration, H.~Abramowicz {\em et.~al.}, {\it {Search for
  single-top production in $ep$ collisions at HERA}},  {\em Phys. Lett.} {\bf
  B708} (2012) 27--36, [\href{http://arxiv.org/abs/1111.3901}{{\tt
  1111.3901}}].

\bibitem{Abdallah:2003wf}
{\bf DELPHI} Collaboration, J.~Abdallah {\em et.~al.}, {\it {Search for single
  top production via FCNC at LEP at $\sqrt{s}$ = 189-GeV to 208-GeV}},  {\em
  Phys. Lett.} {\bf B590} (2004) 21--34,
  [\href{http://arxiv.org/abs/hep-ex/0404014}{{\tt hep-ex/0404014}}].

\bibitem{Sirunyan:2017kkr}
{\bf CMS} Collaboration, A.~M. Sirunyan {\em et.~al.}, {\it {Search for
  associated production of a Z boson with a single top quark and for tZ
  flavour-changing interactions in pp collisions at $ \sqrt{s}=8 $ TeV}},  {\em
  JHEP} {\bf 07} (2017) 003, [\href{http://arxiv.org/abs/1702.01404}{{\tt
  1702.01404}}].

\bibitem{Malekhosseini:2018fgp}
M.~Malekhosseini, M.~Ghominejad, H.~Khanpour, and M.~Mohammadi~Najafabadi, {\it
  {Constraining top quark flavor violation and dipole moments through three and
  four-top quark productions at the LHC}},  {\em Phys. Rev. D} {\bf 98} (2018),
  no.~9 095001, [\href{http://arxiv.org/abs/1804.05598}{{\tt 1804.05598}}].

\bibitem{CMS-DP-2016-064}
{\bf CMS Collaboration} Collaboration, {\it {Updates on Projections of Physics
  Reach with the Upgraded CMS Detector for High Luminosity LHC}}, .

\bibitem{CMS-PAS-FTR-13-016}
{\bf CMS Collaboration} Collaboration, {\it {Projections for Top FCNC Searches
  in 3000/fb at the LHC}},  tech. rep., CERN, Geneva, 2013.

\bibitem{Abada:2019lih}
{\bf FCC} Collaboration, A.~Abada {\em et.~al.}, {\it {FCC Physics
  Opportunities}},  {\em Eur. Phys. J.} {\bf C79} (2019), no.~6 474.

\bibitem{Abada:2019zxq}
{\bf FCC} Collaboration, A.~Abada {\em et.~al.}, {\it {FCC-ee: The Lepton
  Collider}},  {\em Eur. Phys. J. ST} {\bf 228} (2019), no.~2 261--623.

\bibitem{CEPC-SPPCStudyGroup:2015csa}
M.~Ahmad {\em et.~al.}, {\it {CEPC-SPPC Preliminary Conceptual Design Report.
  1. Physics and Detector}}, .

\bibitem{Aihara:2019gcq}
{\bf ILC} Collaboration, H.~Aihara {\em et.~al.}, {\it {The International
  Linear Collider. A Global Project}},
  \href{http://arxiv.org/abs/1901.09829}{{\tt 1901.09829}}.

\bibitem{Baer:2013cma}
{\it {The International Linear Collider Technical Design Report - Volume 2:
  Physics}},  \href{http://arxiv.org/abs/1306.6352}{{\tt 1306.6352}}.

\bibitem{Behnke:2013lya}
H.~Abramowicz {\em et.~al.}, {\it {The International Linear Collider Technical
  Design Report - Volume 4: Detectors}},
  \href{http://arxiv.org/abs/1306.6329}{{\tt 1306.6329}}.

\bibitem{Abramowicz:2013tzc}
{\bf CLIC Detector and Physics Study} Collaboration, H.~Abramowicz {\em
  et.~al.}, {\it {Physics at the CLIC e+e- Linear Collider -- Input to the
  Snowmass process 2013}},  in {\em {Proceedings, 2013 Community Summer Study
  on the Future of U.S. Particle Physics: Snowmass on the Mississippi
  (CSS2013): Minneapolis, MN, USA, July 29-August 6, 2013}}, 2013.
\newblock \href{http://arxiv.org/abs/1307.5288}{{\tt 1307.5288}}.

\bibitem{Brivio:2017vri}
I.~Brivio and M.~Trott, {\it {The Standard Model as an Effective Field
  Theory}},  {\em Phys. Rept.} {\bf 793} (2019) 1--98,
  [\href{http://arxiv.org/abs/1706.08945}{{\tt 1706.08945}}].

\bibitem{Grzadkowski:2010es}
B.~Grzadkowski, M.~Iskrzynski, M.~Misiak, and J.~Rosiek, {\it {Dimension-Six
  Terms in the Standard Model Lagrangian}},  {\em JHEP} {\bf 10} (2010) 085,
  [\href{http://arxiv.org/abs/1008.4884}{{\tt 1008.4884}}].

\bibitem{AguilarSaavedra:2008zc}
J.~A. Aguilar-Saavedra, {\it {A Minimal set of top anomalous couplings}},  {\em
  Nucl. Phys.} {\bf B812} (2009) 181--204,
  [\href{http://arxiv.org/abs/0811.3842}{{\tt 0811.3842}}].

\bibitem{Degrande:2011ua}
C.~Degrande, C.~Duhr, B.~Fuks, D.~Grellscheid, O.~Mattelaer, and T.~Reiter,
  {\it {UFO - The Universal FeynRules Output}},  {\em Comput. Phys. Commun.}
  {\bf 183} (2012) 1201--1214, [\href{http://arxiv.org/abs/1108.2040}{{\tt
  1108.2040}}].

\bibitem{Alwall:2014hca}
J.~Alwall, R.~Frederix, S.~Frixione, V.~Hirschi, F.~Maltoni, O.~Mattelaer,
  H.~S. Shao, T.~Stelzer, P.~Torrielli, and M.~Zaro, {\it {The automated
  computation of tree-level and next-to-leading order differential cross
  sections, and their matching to parton shower simulations}},  {\em JHEP} {\bf
  07} (2014) 079, [\href{http://arxiv.org/abs/1405.0301}{{\tt 1405.0301}}].

\bibitem{Amorim:12}
A.~Amorim, J.~Santiago, N.~Castro, and R.~Santos, {\it {
  http://feynrules.irmp.ucl.ac.be/wiki/GeneralFCNTop}}, .

\bibitem{Alloul:2013bka}
A.~Alloul, N.~D. Christensen, C.~Degrande, C.~Duhr, and B.~Fuks, {\it
  {FeynRules 2.0 - A complete toolbox for tree-level phenomenology}},  {\em
  Comput. Phys. Commun.} {\bf 185} (2014) 2250--2300,
  [\href{http://arxiv.org/abs/1310.1921}{{\tt 1310.1921}}].

\bibitem{Skands:2014pea}
P.~Skands, S.~Carrazza, and J.~Rojo, {\it {Tuning PYTHIA 8.1: the Monash 2013
  Tune}},  {\em Eur. Phys. J.} {\bf C74} (2014), no.~8 3024,
  [\href{http://arxiv.org/abs/1404.5630}{{\tt 1404.5630}}].

\bibitem{deFavereau:2013fsa}
{\bf DELPHES 3} Collaboration, J.~de~Favereau, C.~Delaere, P.~Demin,
  A.~Giammanco, V.~Lemaitre, A.~Mertens, and M.~Selvaggi, {\it {DELPHES 3, A
  modular framework for fast simulation of a generic collider experiment}},
  {\em JHEP} {\bf 02} (2014) 057, [\href{http://arxiv.org/abs/1307.6346}{{\tt
  1307.6346}}].

\bibitem{Tanabashi:2018oca}
{\bf Particle Data Group} Collaboration, M.~Tanabashi {\em et.~al.}, {\it
  {Review of Particle Physics}},  {\em Phys. Rev. D} {\bf 98} (2018), no.~3
  030001.

\bibitem{Backovic:2013bga}
M.~Backovic, O.~Gabizon, J.~Juknevich, G.~Perez, and Y.~Soreq, {\it {Measuring
  boosted tops in semi-leptonic $t\bar t$ events for the standard model and
  beyond}},  {\em JHEP} {\bf 04} (2014) 176,
  [\href{http://arxiv.org/abs/1311.2962}{{\tt 1311.2962}}].

\bibitem{MoortgatPick:2005cw}
G.~Moortgat-Pick {\em et.~al.}, {\it {The Role of polarized positrons and
  electrons in revealing fundamental interactions at the linear collider}},
  {\em Phys. Rept.} {\bf 460} (2008) 131--243,
  [\href{http://arxiv.org/abs/hep-ph/0507011}{{\tt hep-ph/0507011}}].

\end{thebibliography}\endgroup

\end{document}